\documentclass[%
 reprint,
superscriptaddress,
floatfix,
 amsmath,amssymb,
 aps,
 pre,
]{revtex4-2}

\usepackage{graphicx}
\usepackage{dcolumn}
\usepackage{bm}
\usepackage{xparse}
\usepackage{xcolor}

\makeatletter
\DeclareDocumentCommand\quantity{}{{\ifnum\z@=`}\fi\@quantity}
\DeclareDocumentCommand\@quantity{ t\big t\Big t\bigg t\Bigg g o d() d|| }
{ 
	\IfBooleanTF{#1}{\let\ltag\bigl \let\rtag\bigr}{
		\IfBooleanTF{#2}{\let\ltag\Bigl \let\rtag\Bigr}{
			\IfBooleanTF{#3}{\let\ltag\biggl \let\rtag\biggr}{
				\IfBooleanTF{#4}
				{\let\ltag\Biggl \let\rtag\Biggr}
				{\let\ltag\left \let\rtag\right}
			}
		}
	}
	\IfNoValueTF{#5}{
		\IfNoValueTF{#6}{
			\IfNoValueTF{#7}{
				\IfNoValueTF{#8}
				{()}
				{\ltag\lvert{#8}\rtag\rvert}
			}
			{\ltag(#7\rtag) \IfNoValueTF{#8}{}{|#8|}}
		}
		{\ltag[#6\rtag] \IfNoValueTF{#7}{}{(#7)} \IfNoValueTF{#8}{}{|#8|}}
	}
	{\ltag\lbrace#5\rtag\rbrace  \IfNoValueTF{#6}{}{[#6]} \IfNoValueTF{#7}{}{(#7)} \IfNoValueTF{#8}{}{|#8|}}
	\ifnum\z@=`{\fi}
}
\DeclareDocumentCommand\qty{}{\quantity} 
\makeatother

\newcommand{\kb}{k_{\textrm{B}}}
\newcommand{\Gg}{\bm{G}\cdot\bm{g}}
\newcommand{\Ggh}{\bm{G}\cdot\hat{\bm{g}}}

\begin{document}

\title{Kinetic Theory for Electronic Transport Properties of Warm Dense Matter: Chapman-Enskog Solution of the Uehling-Uhlenbeck Equation}

\author{Lucas J. Babati}
\affiliation{Nuclear Engineering and Radiological Sciences, University of Michigan, Ann Arbor, Michigan, 48109, USA}%

\author{Nathaniel R. Shaffer}
\affiliation{Laboratory for Laser Energetics, University of Rochester, Rochester, New York, 14623, USA}%

\author{Louis Jose}
\affiliation{Nuclear Engineering and Radiological Sciences, University of Michigan, Ann Arbor, Michigan, 48109, USA}%

\author{Scott D. Baalrud}
\affiliation{Nuclear Engineering and Radiological Sciences, University of Michigan, Ann Arbor, Michigan, 48109, USA}%
\email{baalrud@umich.edu}

\date{\today}

\begin{abstract}
A kinetic theory is developed to describe the electrical conductivity, thermal conductivity,
and electrothermal coefficients in warm dense plasmas. It models electron degeneracy
using the Uehling-Uhlenbeck equation, diffraction by computing scattering
cross sections quantum mechanically, and strong coupling by treating the scattering
events using the potential of mean force. A key advancement detailed here is the development of 
a Chapman-Enskog solution of the Uehling-Uhlenbeck equation for hydrodynamic
transport coefficients. The result is a model
which accurately predicts transport coefficients spanning from warm dense matter conditions
through hot dilute plasmas, including the influence of electron-electron interactions.
Results are compared with quantum molecular dynamics simulations, experiments, and
other models. The present method is able to capture the ``Spitzer'' terms in the classical
plasma limit, while also capturing the correct degenerate limit. The transition between
these limits in the warm dense matter regime is explained in terms of the availability
of states for electron scattering.
\end{abstract}

\maketitle

\section{Introduction}
\label{sec:intro}
Plasmas are commonly modeled using hydrodynamics, including in
the warm dense matter regime. For example, experiments on the National Ignition Facility (NIF)~\cite{NIF}, the Z machine~\cite{GomezPRL2014}, and astrophysical objects like
white dwarf stars~\cite{SaumonPhysRep2022} and the interior of giant planets~\cite{MilitzerJGR2016} reach warm dense matter conditions.
Hydrodynamics models require a description of the transport coefficients and equation
of state as a function of the mass density, temperature, and species concentrations as an input. 
In plasmas, these are traditionally supplied by the Chapman-Enskog solution of the Boltzmann equation~\cite{FK, CC}. 
In condensed systems, they are often computed from Density Functional Theory Molecular Dynamics (DFT-MD)~\cite{LenoskyPRB1997,AlfePRL1998,HolstPRB2011}. 
However, neither approach applies to warm dense matter. 
Traditional plasma models break down because they do not treat the strong Coulomb
correlations of ions or the quantum degeneracy of electrons that define the warm dense matter regime. 
DFT-MD methods break down because they become too computationally expensive at high
temperatures to practically compute transport properties~\cite{BlanchetPOP2020}, and because they do not
adequately account for electron-electron interactions~\cite{FrenchPRE2022,BergermannPOP2026}, which we show are important in warm dense matter. 

Here, a new approach is developed to extend plasma kinetic theory into the 
warm dense matter regime. 
It models strong correlation effects by treating interactions using the potential of 
mean force~\cite{HansenMcD, StarrettPRE2013, StarrettHEDP2017, ShafferPRE2020}, Pauli
blocking using the Boltzmann-Uehling-Uhlenbeck (BUU) equation~\cite{UehlingPR1933,
UehlingPR1934}, and electron diffraction by computing scattering cross sections quantum mechanically~\cite{Sakurai}. 
Hydrodynamic transport properties are computed from a Chapman-Enskog expansion of the BUU equation~\cite{UehlingPR1933, UehlingPR1934}. 
The result is a convergent framework to determine transport coefficients ranging
from the classical plasma limit through the warm dense matter regime. This 
theory can be viewed as the semiclassical generalization of mean force kinetic 
theory~\cite{BaalrudPRL2013, BaalrudPOP2019}.

\begin{figure}
    \includegraphics[width=0.48\textwidth]{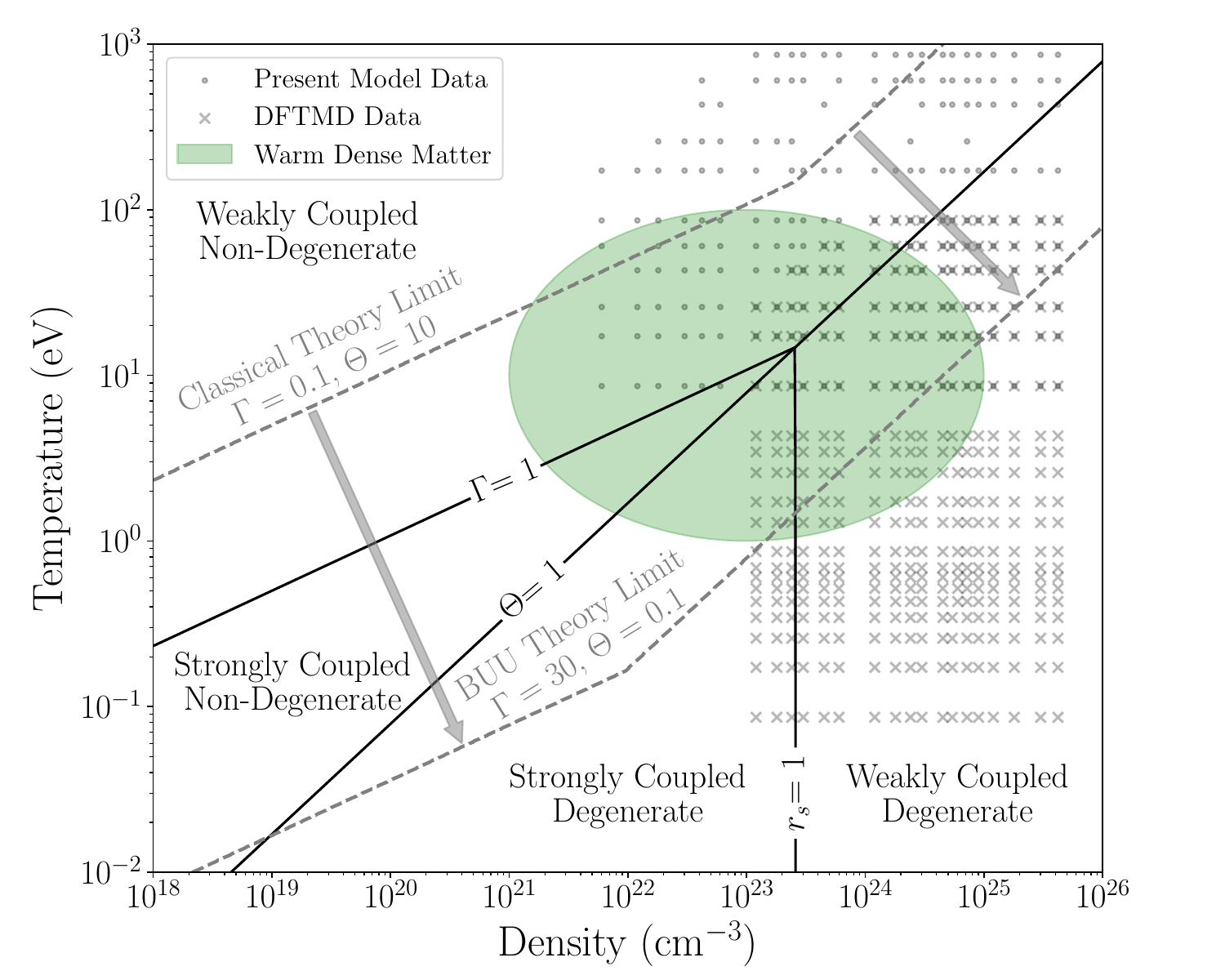}
    \caption{\label{fig:phase_plot} Warm dense matter (denoted in green) is defined as the region of density-temperature conditions where the dimensionless
    quantities $\Gamma$, $\Theta$, and $r_s$ are near unity.  
    It is expected that traditional kinetic theories work best in 
    the weakly coupled degenerate region up to the grey dotted line marked ``Classical 
    Theory Limit", whereas the present model extends this limit through the warm 
    dense matter regime to the dotted line marked ``BUU Theory Limit".}
\end{figure}

Warm dense matter is typically defined as a state of matter where the system's 
average kinetic, potential, and Fermi energies are all roughly equal.
The regions in density and temperature space are mapped out in Fig.~\ref{fig:phase_plot}, 
where the solid lines correspond to transition points where the various ratios between these 
energies are unity. 
The first ratio is the Coulomb coupling parameter
\begin{equation}
    \Gamma_{ij} = \frac{q_i q_j}{4\pi\epsilon_0 a \kb T}
\end{equation}
where $q_i$ is the charge of species $i$, $a=(3/4 \pi n)^{1/3}$ is the average 
interatomic spacing, or the Wigner-Seitz radius, and $T$ is the temperature. 
This parameter measures the ratio between the average potential and kinetic energies.  
If $\Gamma \ll 1$ the system is 
weakly coupled, whereas if $\Gamma \gg 1$ it is strongly coupled. 
Another energy comparison is the degeneracy parameter,
\begin{equation}
    \Theta = \frac{\kb T}{E_{\textrm{F}}}
\end{equation}
where $E_{\textrm{F}} = \hbar^2 (3\pi^2 n_e)^{2/3}/(2m_e)$ is the Fermi energy 
of the electrons in the system. If $\Theta \gg 1$ electrons in the 
system behave classically, whereas if $\Theta \ll 1$ quantum effects like Pauli blocking and diffraction
become dominating effects.
Further in this regime, the kinetic energy of electrons is determined by the Fermi energy rather than 
the thermal energy. Due to this, the Coulomb coupling parameter is redefined for electrons as 
\begin{equation}
    r_s = \frac{a}{a_0} = 1.8 \frac{e^2}{4\pi\epsilon_0 a E_{\textrm{F}}}
\end{equation}
where $a_0 = 4\pi\epsilon_0 \hbar^2 / m_e e^2$ is the Bohr radius. This parameter 
has the same meaning as $\Gamma$, but applies in the degenerate regime. 
Warm dense matter corresponds to the conditions where all of these parameters are 
roughly unity, which makes developing a theory for the system challenging, as none
of these dimensionless parameters can be used as an expansion parameter.

Transport properties of warm dense matter are typically  computed using DFT-MD, or by generalizing plasma 
kinetic theories. 
DFT-MD comes from 
condensed matter theory~\cite{HohenbergPR1964, KohnPR1965, MerminPR1965}, corresponding
to the dense and low temperature conditions toward the bottom right of Fig.~\ref{fig:phase_plot}.
An example of the conditions where DFT-MD can be used is indicated by the grey crosses,
which show parameters of recent simulations of hydrogen plasmas~\cite{BergermannPOP2026}
that push the limits of what is computationally possible with DFT-MD. As the temperature increases, or
density decreases, the computational cost of DFT-MD rises steeply~\cite{BlanchetPOP2020}. Despite this, much research continues to extend DFT-MD further into 
warm dense matter conditions~\cite{WhitePRL2020} and many studies of electrical transport 
coefficients have been conducted~\cite{RedmerPhysRep1997, HolstPRB2011, WittePOP2018, BethkenhagenPRR2020, FrenchPRE2022, StanekPOP2024, BergermannPOP2026, Ramakrishna, SharmaPOP2026}.
DFT-MD is usually considered the most accurate method for computing electronic 
transport coefficients, achieved at the cost of extreme computational
expense. However, DFT-MD is a mean field theory, and cannot capture electron-electron
interactions~\cite{FrenchPRE2022}. Although these are expected to be negligible in condensed matter systems
due to Pauli blocking, they are known to be important in a weakly coupled plasma,
determining a dominate contribution to electrical and thermal conductivity~\cite{SpitzerPhysRev1953}. 
It is an open question to describe the role of electron-electron interactions in the
warm dense matter regime, and to bridge this gap with the traditional plasma limit. 

Traditional plasma kinetic theories~\cite{Boltzmann1970,Landau1936,Gradhand1958,LenardAnnPhys1960,BalescuPhysFluids1960,GuernseyPhysFluids1964,FK,Harris1971,LeeMore}
apply in classical weakly coupled systems, corresponding to the lower density higher temperature conditions in the top left 
of Fig.~\ref{fig:phase_plot}. 
Although the kinetic equations can be written in different forms, they are all based on an expansion for $\Gamma \ll 1$ and $\Theta \gg 1$. 
There has 
been significant effort to relax these approximations in an effort to extend plasma kinetic 
theory to strongly coupled and degenerate systems~\cite{GouldPhysRev1967, LampePR1968,
WilliamsPhysFluids1969, BoerckerAOP1979, RopkePRA1988, Bonitz, RopkePOP2026}. However, none have
yet demonstrated the ability to predict electronic transport properties spanning the
warm dense matter regime. Some have addressed strong coupling in a classical statistical
plasma~\cite{PaquetteAPJS1986, BaalrudPRL2013, StantonPRE2016}. Others have addressed
quantum electron interactions in weakly coupled plasmas~\cite{DaligaultPOP2016,DaligaultPOP2018}. 
Further, some have also attempted to combine aspects of strong ion coupling and electron
degeneracy through model relaxation time approximations~\cite{StarrettHEDP2017,FrenchPRE2022}. What is still lacking is
a kinetic theory for transport in warm dense matter based on systematic approximations. 

Here, we develop a kinetic theory for warm dense matter based on a combination of
the mean force kinetic theory concept, and the BUU equation. Mean force kinetic theory
is a Boltzmann-like kinetic equation, but where binary interactions occur via the
potential of mean force rather than the bare force~\cite{BaalrudPRL2013,BaalrudPOP2019}. 
It has been previously shown to extend plasma kinetic theory in classical systems
to $\Gamma \lesssim 20$~\cite{BaalrudPRL2013,BaalrudPOP2019}. When used in combination
with an average atom model to address screening by degenerate electrons, it was also
shown to capture ion transport properties in warm dense matter~\cite{DaligaultPRL2016}. 
Here, we extend this to electronic transport by treating degenerate electron statistics
with the BUU equation, and electron diffraction in scattering events by computing
the cross sections quantum mechanically~\cite{Sakurai}. Although, these steps have been applied for
ion-electron interactions previously~\cite{RightleyPRE2021,BabatiPRE2026}, only a
semiclassical solution was obtained because the ions are treated classically. 
Therefore, the role of electron-electron interactions was not addressed. 

Perhaps the closest previous work to the present method is Ref.~\cite{ShafferPRE2020_CE},
which used the mean force concept for interactions, but treated electron-electron
interactions using the quantum-Landau-Fokker-Plank (qLFP) equation from Daligault~\cite{DaligaultPOP2016, DaligaultPOP2018}. 
The present model aims to improve on this theory, as the qLFP equation corresponds to
a small scattering angle expansion of the BUU equation~\cite{DaligaultPOP2016}. 
Here, the BUU equation is solved with a Chapman-Enskog expansion~\cite{Enskog1917, Chapman1916, Chapman1918, Braginskii1965, CC, FK, HCB}   
and no small scattering angle approximation is taken. This expansion has been done 
before, but was limited to a nondegenerate system~\cite{UehlingPR1934} ($\Theta \gg 1$),
or neutral gases~\cite{WuJCP2019}.
The present work fills the gap between plasma kinetic theory and DFT-MD
by extending plasma kinetic theory into the warm dense matter regime. 
This can be seen in Fig.~\ref{fig:phase_plot} as the gap between the grey 
dotted lines labeled ``Classical Theory Limit'' and ``BUU Theory Limit''. Also,
the dots on the figure represent a dataset which was run with the present method, 
and show that it can access conditions within warm dense matter typically limited 
to DFT-MD, but also extends to the weakly coupled plasma regime.

A companion paper describes the main results of the calculations and how they properly
extend previous results to address electron-electron interactions contributions~\cite{BabatiPRL2026}. 
Here, the major calculational challenge of developing a Chapman-Enskog solution of
the BUU equation is described. Some further results are also presented.

The paper is outlined as follows. In Section~\ref{sec:pmf}, the potential of mean 
force and classical mean force kinetic theory are introduced. In Section~\ref{sec:BUUeqn},
the Boltzmann-Uehling-Uhlenbeck equation is introduced and its properties are 
discussed. In Section~\ref{sec:CE}, the Chapman-Enskog expansion is performed on 
the linearized BUU equation. Section~\ref{sec:var} discusses the variational 
procedure to obtain transport coefficients and Section~\ref{sec:electrons} gives
explicit expressions for electronic transport coefficients. In Section~\ref{sec:results},
electronic transport coefficients are calculated for various elements, compared to
DFT-MD and experimental data where it exists, and their physical significance is 
discussed.

\section{Potential of Mean Force}
\label{sec:pmf} 
Classical kinetic theories are often derived based on closure approximations to the BBGKY 
hierarchy. For example, the Boltzmann equation can be derived assuming that 
correlations between 3-body and higher interactions are negligible. 
Formally, this corresponds to taking the three body distribution function to vanish, $f^{(3)}=0$;
an approximation that can be justified in the limit of asymptotically small coupling strength $\Gamma \ll 1$. 
If the system is expanded 
in terms of deviations from equilibrium instead of the strength of correlations, the hierarchy can be closed by assuming
that 3-body and higher interactions are at equilibrium. 
Specifically, mean force 
kinetic theory~\cite{BaalrudPOP2019} is derived from the closure $\Delta f^{(3)}=0$,
where $\Delta f^{(3)} = f^{(3)}-f^{(2)}f_o^{(3)}/f_o^{(2)}$ and $f^{(n)}_o$ is the equilibrium $n$-particle distribution function. 
This preserves the exact equilibrium state at all orders of the hierarchy and allows the theory to extend to higher coupling strength $\Gamma \lesssim 20$. 
It leads to a Boltzmann-like kinetic theory, but 
where the interactions are mediated by the potential of mean force~\cite{HansenMcD}
instead of the Coulomb potential. 

\begin{figure*}
    \includegraphics[width=0.98\textwidth]{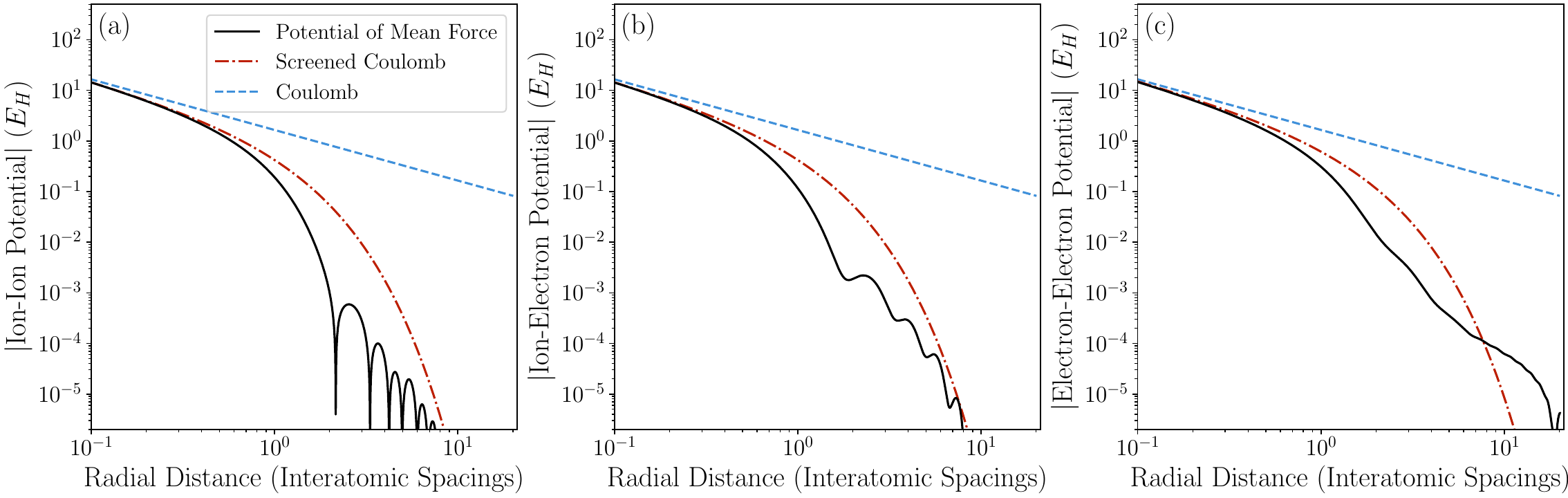}
    \caption{\label{fig:pmf} Absolute value of the (a) ion-ion~\cite{StarrettPRE2013}, (b) ion-electron~\cite{StarrettHEDP2017}, 
        and (c) electron-electron~\cite{ShafferPRE2020} potentials of mean force plotted as a 
        function of radial distance from a central particle for a hydrogen plasma at 17.2~eV and 12~g/cm$^{3}$ 
        ($\Gamma =2.6$, $\Theta=0.13$). The screened Coulomb potential uses 
        the screening length described in Ref~\cite{StantonPOP2021}. Note that the potential of mean force 
        maintains the correct screening, while also capturing correlations between 
    other particles. Potential energy is given in Hartrees ($E_\textrm{H}$), where 1~Hartree $\approx27.2$~eV.}
\end{figure*}

In classical statistical systems, the potential of mean force is the potential between two particles 
held at fixed positions while the rest of the $N-2$ particles are thermodynamically
averaged over. 
It is related to the radial distribution function as $\phi_{ij} = -k_\textrm{B}T \ln [g_{ij}(r)]$~\cite{HansenMcD}. 
It is commonly calculated using the Ornstein-Zernicke equation~\cite{Ornstein1914},
and for Coulombic systems is closed with the hypernetted chain closure~\cite{HansenMcD}.
The hypernetted chain closure results from the assumption that the difference 
between direct and indirect correlations is small, which is a good approximation 
for $\Gamma \lesssim 100$~\cite{BaalrudPOP2014}. 

For quantum systems, the potential of mean force cannot be defined in the same 
way. This is because momenta and positions of particles cannot be factored and 
independently averaged over. Nevertheless, an analog of the potential of mean force has 
been defined for warm dense matter through the idea of an average atom~\cite{StarrettPRE2013}. The average
atom is an atom with a classical point nucleus of charge $Z$ and electronic structure 
determined by either density functional theory on a set of hydrogenic wave functions~\cite{StarrettPRE2013},
or semiclassically through a Thomas-Fermi approximation~\cite{StarrettHEDP2014}.
They represent the average electronic structure of all 
atoms of a particular species in a plasma. From this picture, an average ionization 
of the system, $\bar{Z}$, can be defined as the number of electrons with 
positive energy. The full rigorous definition used in this work is from Eq.~(73) of Ref.~\cite{StarrettPRE2013}. This is 
one choice for the average ionization that is convenient for the average atom model, but 
in general different models may use different definitions. In the kinetic theory, 
$\bar{Z}$ determines the free electron population that influences collisional transport 
in the system. This means that only elastic scattering will be considered in this work. 

To access interaction potentials, the average atoms are then coupled with a background plasma so that correlation
functions can be found from a quantum Ornstein-Zernicke equation and hypernetted chain closure.
Explicit procedures for producing each potential relevant to a plasma are explained in Ref.~\cite{StarrettPRE2013}
for the ion-ion interaction, Ref.~\cite{StarrettHEDP2017} for the ion-electron 
interaction, and Ref.~\cite{ShafferPRE2020} for the electron-electron interaction.
All of these works use the same average atom model as the basis for the calculation, making the
set of potentials obtained consistent with each other. These potentials of mean force are used in the quantum kinetic 
theory described below without rigorous proof, but motivated by classical mean force kinetic theory.

A limitation of the average atom model is that it cannot capture bonding. This is because, by 
definition, only a single atom's electronic structure is solved for and the 
interaction between other atoms is only considered in a spherically averaged sense. Thus, 
if the system is cold and dilute, for example ambient air, the average atom approximation 
is insufficient. It is best in dense high temperature systems, such as warm dense matter. 
Further, when the average atoms are coupled to the background plasma, the free electrons are 
treated as an ideal Fermi gas. In certain systems,
the free electron density of states can feature non-ideal characteristics~\cite{KinneyPRA2026}
which would not be captured in these calculations. This can be significant in 
extremely dense systems where electrons are pressure ionized but still retain 
a hint of electronic structure.

Despite these limitations, these potentials have been shown to capture the most important correlation
physics in warm dense matter~\cite{StarrettPRE2013, StarrettHEDP2017, ShafferPRE2020}
and allow kinetic theories to extend to stronger coupling and higher degeneracy~\cite{DaligaultPRL2016, ShafferPRE2020_CE, RightleyPRE2021, BabatiPRE2026}.
An example of this can be seen in Fig.~\ref{fig:pmf}, where the potentials of 
mean force are plotted for a hydrogen plasma at 5~eV and 1.67~g/cm$^3$. In a classical 
plasma, the potential of mean force is only a function of the Coulomb coupling parameter, $\Gamma$.
So, for a fully ionized hydrogen plasma, like the example shown, electrons and ions
would be treated on equal footing and the ion-ion and electron-electron potentials
of mean force would be identical. As can be seen in Fig.~\ref{fig:pmf}, when the electrons
become degenerate, the potential of mean force for each interaction differs substantially. 
This is because degeneracy influences the way in which electrons screen the ions and themselves. 
To lowest order, this is illustrated by how Debye screening differs from Thomas-Fermi screening in weakly coupled systems.  
This can be seen in Fig.~\ref{fig:pmf},
where the screening length used for the ion-ion and ion-electron potentials in the``screened Coulomb'' 
result is provided from a model with an effective screening length that asymptotes to Debye or Thomas-Fermi in the appropriate limits~\cite{StantonPRE2016, StantonPOP2021}. 
Clearly, the screened Coulomb potential differs from the full 
potential of mean force computed from the average atom model. 
This is due to the strong spatial correlations of particles at these conditions, which are included in the two-component average atom model, but are absent in a simple screened Coulomb potential. 
The strong correlations cause oscillations in the potentials that significantly influence the effective interaction force and ultimate transport rates that are calculated from them. 
Oscillations from strong correlations are largest in the ion-ion interaction because ions have classical statistics, they are weaker in the ion-electron interaction because these interactions are semiclassical (classical ions, degenerate electrons), and weakest in the electron-electron interaction because they are degenerate. 
Pauli blocking effectively weakens correlations, which influences the potential of mean force. 
These potentials of mean force supply the input to the kinetic theory. 


\section{Boltzmann-Uehling-Uhlenbeck Equation}
\label{sec:BUUeqn}
The Boltzmann-Uehling-Uhlenbeck (BUU) equation~\cite{UehlingPR1933, UehlingPR1934} is a
semiclassical generalization of the classical Boltzmann equation. For a system 
of $K$ species, it is given by 
\begin{equation}
    \frac{\partial f_i}{\partial t} 
    + \frac{\bm{p}_i}{m_i}\cdot \frac{\partial f_i}{\partial \bm{r}} 
    + \bm{F}_i\cdot \frac{\partial f_i}{\partial \bm{p}} = \sum_{j=1}^{K}\mathcal{C}\qty(f_i,f_j).
    \label{eqn:BUU}
\end{equation}
Here, $i$ and $j$ index species in the system, $f_i(\bm{r}, \bm{p}, t)$ is the Wigner function
for species $i$, $\bm{p}$ is the momentum, $m$ is the mass, $\bm{F}$ is the external force, $\partial/\partial \bm{r}$ is a spatial 
gradient, and $\partial/\partial \bm{p}$ is the gradient in momentum space. The collision term is given by
\begin{align}
    \mathcal{C}\qty(f_i,f_j) = \int d^3p_j& d\Omega \frac{d \sigma}{d \Omega} u \Big[\hat{f}_i\hat{f}_j\qty(1+\delta_i\theta_i f_i)\qty(1+\delta_j\theta_j f_j) \nonumber \\ 
                                          &-f_if_j\qty(1+\delta_i\theta_i \hat{f}_i)\qty(1+\delta_j\theta_j \hat{f}_j)\Big]
                                          \label{eqn:BUU_operator}
\end{align}
where hatted quantities are taken post collision and unhatted quantities are taken
pre collision. Here $d\sigma/d\Omega$ is the differential scattering cross section,
$u= \qty|\bm{p}_i/m_i - \bm{p}_j/m_j|$ is the relative velocity of a collision, $\delta_i = -1, 0, 1$
for Fermi-Dirac, Boltzmann, and Bose-Einstein statistics respectively, $\theta_i = (2\pi\hbar)^3/s_i$
is the phase space volume per particle, and $s_i$ is the spin multiplicity.
If the system is assumed to be classical, then $\delta_i = \delta_j = 0$ and the 
classical Boltzmann collision operator is recovered. 

Similar to the process of obtaining the classical Boltzmann equation via the 
Liouville equation and the BBGKY hierarchy, Eq.~(\ref{eqn:BUU}) can be found from 
the equation of motion of the density matrix operator of quantum statistical mechanics~\cite{Bonitz, BoerckerAOP1979}.
Due to this, Eq.~(\ref{eqn:BUU}) describes the evolution of the Wigner function, $f$, 
which is a representation
of the density matrix itself. In the classical limit it becomes the velocity distribution function. To obtain this form, a weak coupling approximation 
akin to molecular chaos must be made. This means Eq.~(\ref{eqn:BUU_operator}) only 
describes the interactions of pairs of statistically independent particles. 
As with the classical mean force kinetic theory, the aspect of modeling strong coupling comes from treating binary interactions as occurring through the potential of mean force. 

\subsection{Properties}
\label{subsec:props} 
The BUU equation has many properties that are shared with the Boltzmann equation, but
others which are unique. The main differences occur in the factors $(1+\delta\theta f)$,
which depending on the statistics, can include Pauli blocking for fermions or 
condensation for bosons. In particular, these factors lead the equilibrium distribution 
to be a Fermi-Dirac or Bose-Einstein distribution defined as 
\begin{equation}
    f_i^{(0)}\qty(\bm{p}) = \frac{1}{\theta_i} \frac{1}{\exp\qty(\frac{\beta(\bm{p}-m_i\bm{v})^2}{2m_i}-\beta\mu_i) - \delta_i}.
    \label{eqn:FD}
\end{equation}
Here, $\bm{v}$ is the average flow speed of the system, $\beta = 1/\kb T$ is the 
inverse temperature, and $\mu_i$ is the chemical potential of species $i$. The 
chemical potential is uniquely defined using the normalization condition $n_i = \int d^3 p f_i^{(0)}(\bm{p})$, which 
gives
\begin{equation}
    \mathcal{Q}_{1/2}\qty(\beta \mu) = \frac{4}{3\sqrt{\pi}}\Theta^{-3/2}.
\end{equation}
The integrals $\mathcal{Q}_{n}$ are Fermi-Dirac (or Bose-Einstein)
integrals defined as
\begin{equation}
    \mathcal{Q}_n(x) = \frac{1}{\Gamma(n+1)}\int \frac{y^n}{e^{y-x} - \delta} dy,
    \label{eqn:fd_int}
\end{equation}
where $\Gamma(n)$ is the gamma function.
The equilibrium distribution function has the unique property that if $f_i$ and 
$f_j$ are at equilibrium defined by Eq.~(\ref{eqn:FD}), then $\mathcal{C}\qty(f_i, f_j) = 0$~\cite{UehlingPR1933,DaligaultPOP2016,DaligaultPOP2018}.

Despite these differences, the BUU equation shares other properties that are similar
to the Boltzmann equation. In particular, it only considers elastic scattering processes.
This means that mass (or particle number), momentum, and 
kinetic energy are all conserved. Mathematically, this can be expressed as moments of the collision 
operator for $K$ species
\begin{align}
    &\int d^3 p_i \mathcal{C}(f_i, f_j) = 0, \\
    &\sum_{i,j=1}^K \int d^3 p_i \bm{p}_i \mathcal{C}(f_i, f_j) = 0, \\
    &\sum_{i,j=1}^K \int d^3 p_i \frac{p_i^2}{2m_i} \mathcal{C}(f_i, f_j) = 0.
\end{align}
The set of these conserved quantities, the mass of each species,
the total momentum in each direction, and the total kinetic energy of the system,
are unique quantities and will be referred to as collisional or summational invariants.

\subsection{Linearized BUU Equation}
\label{subsec:lin}
Paramount to the Chapman-Enskog procedure is the existence of 
a linearized collision operator. This can be found from the BUU operator, Eq.~(\ref{eqn:BUU_operator}),
by perturbing the equilibrium distribution. Let $f^{(0)}_i$ be the equilibrium distribution,
and $f^{(1)}_i$ be the perturbation, then the linearized collision operator becomes
\begin{align}
    \mathcal{J}\qty(f_i, f_j) = \int &d^3p_j d\Omega \frac{d \sigma}{d \Omega} u h_i h_j  \Big[\hat{\chi}_i + \hat{\chi}_j - \chi_i - \chi_j \Big],
                                          \label{eqn:BUU_lin}
\end{align}
where $h_i = f^{(0)}_i\qty(1+\delta_i\theta_i \hat{f}^{(0)}_i)$, $\chi_i = f^{(1)}_i/w_i$,
and $w_i = f^{(0)}_i\qty(1+\delta_i\theta_i f^{(0)}_i)$. Notice the slight difference 
between $h$ and $w$, $w$ only contains pre or post collision momenta, whereas $h$ 
depends on both.

This operator motivates a bilinear operation, known as the bracket integral~\cite{FK}. Let 
$F$ and $G$ be functions of momentum, $\bm{p}$, then for $K$ species the bracket integral is defined
as 
\begin{equation}
    \qty[F,G] = \sum_{i,j=1}^{K} \frac{n_in_j}{n^2} \int d^3 p_i G_i I_{ij}(F).
    \label{eqn:bracket}
\end{equation}
Here, $n$ is the total number density, $n_i$ is the number density of 
species $i$, and $I_{ij}$ is defined as 
\begin{equation}
    I_{ij}(F) = \frac{1}{n_i n_j} \int d^3p_j d\Omega \frac{d \sigma}{d \Omega} u h_i h_j  \Big[F_i + F_j - \hat{F}_i - \hat{F}_j \Big].
    \label{eqn:I_int}
\end{equation}
The linearized collision operator is related to this $I$ operator
by, 
\begin{equation}
    \mathcal{J}(f_i, f_j) = -n_in_jI_{ij}(\chi).
    \label{eqn:lin_To_I}
\end{equation}
To make computations simpler, the integral operator $I_{ij}$ can be broken down into two pieces defined as 
\begin{subequations}
\begin{align}
    I_{ij,i}(F) &= \frac{1}{n_i n_j} \int d^3p_j d\Omega \frac{d \sigma}{d \Omega} u h_i h_j  \Big[F_i - \hat{F}_i\Big],
\end{align}
and
\begin{align}
    I_{ij,j}(F) &= \frac{1}{n_i n_j} \int d^3p_j d\Omega \frac{d \sigma}{d \Omega} u h_i h_j  \Big[F_j - \hat{F}_j\Big].
\end{align}
\end{subequations}
With these definitions, the bracket integral itself can be broken into pieces. Defining
\begin{subequations}
\begin{equation}
    [F, G]^\prime_{ij} = \int d^3 p_i G_i I_{ij,i}(F),
    \label{eqn:brack_p}
\end{equation}
and 
\begin{equation}
    [F, G]^{\prime\prime}_{ij} = \int d^3 p_i G_i I_{ij,j}(F),
    \label{eqn:brack_pp}
\end{equation}
\end{subequations}
the total bracket integral from Eq.~(\ref{eqn:bracket}) is 
\begin{equation}
\label{eq:FG1}
    [F,G] = \sum_{i,j=0}^{K} \frac{n_i n_j}{n^2} \qty([F, G]_{ij}^{\prime} + [F, G]_{ij}^{\prime\prime}).
\end{equation}
This form of the bracket integral will prove useful in the Chapman-Enskog 
expansion.

Due to symmetries within the 
differential scattering cross section, specifically that elastic collisions obey the principle
of detailed balance~\cite{Boltzmann1970, Reif1965, Cercignani1969, FK, Sakurai}, Eq.~(\ref{eq:FG1}) can
be written in the form
\begin{align}
    \qty[F,G] = &\frac{1}{4n^2} \sum_{i,j}  \int d^3 p_i d^3 p_j d\Omega \frac{d\sigma}{d\Omega} u h_i h_j \nonumber \\ 
    \times&\Big[F_i + F_j - \hat{F}_i - \hat{F}_j \Big] \Big[G_i + G_j - \hat{G}_i - \hat{G}_j \Big].
    \label{eqn:bracket_other}
\end{align}
It is clear to see then that the bracket integral is symmetric about its arguments, $[F,G] = [G,F]$.    
In practice, the symmetric form of Eq.~(\ref{eqn:bracket_other}) is more 
useful than Eq.~(\ref{eqn:bracket}) for the evaluation of transport coefficients.

So far the arguments of the bracket integral, $F$ and $G$, have been left ambiguous. These may be of any tensoral rank, as long as they share the same rank and the appropriate operation is used to take the scalar
product. In other words, if the product was expressed in index notation, all indicies
must be summed over and reduced. 

\section{Chapman-Enskog Expansion}
\label{sec:CE} 
The Chapman-Enskog procedure is a well documented process, as described in
Ref.~\cite{FK} or Ref.~\cite{CC}. The procedure presented here will closely 
follow Ref.~\cite{FK}, and is very similar to previous work on the quantum Landau-Fokker-Plank
operator by Daligault~\cite{DaligaultPOP2018}. 
The main distinction here is the use of the linearized BUU equation and the associated form of the bracket integrals. 
From this point forward, the analysis will be specific
to a system of fermions, since the application of interest is a plasma.

\subsection{Distribution Function}
\label{subsec:dist} 
First, it is asserted that the distribution function $f$ contains all thermodynamic
and hydrodynamic information about the system, and that these can be expressed from moments of the 
distribution function. In particular,
\begin{equation}
    n_i = \int d^3 p f_i(\bm{p}_i),
\end{equation}
is the number density of species $i$, and the total number density is
$n = \sum_{i} n_i$. Similarly, the mass density is defined as $\rho = \sum_{i} \rho_i = \sum_{i} m_i n_i$
where $m_i$ is the mass of species $i$. Additionally, the average flow velocity is defined 
as 
\begin{equation}
     \bm{v} = \frac{1}{\rho} \sum_{i=1}^{K} \int d^3 p_i\, \bm{p}_i f_i(\bm{p}_i),
\end{equation}
and the total internal energy per unit mass as 
\begin{equation}
  u = \frac{1}{\rho} \sum_{i=1}^{K} \int d^3 p_i \frac{P_i^2}{2m_i} f_i(\bm{p}_i),
\end{equation}
where $\bm{P}_i = \bm{p}_i - m_i \bm{v}$ is the peculiar momentum, or the 
momentum in the center of mass frame.

In a similar way, thermodynamic fluxes can be defined. Of particular interest are
the diffusion velocity, $\bm{V}_i$, defined as
\begin{equation}
    \rho_i \bm{V}_i = \int d^3 p_i \bm{P}_i f_i(\bm{p}_i)
\end{equation}
the pressure tensor,
\begin{equation}
    \bm{\mathsf{\Pi}} = \sum_{i=1}^{K} \frac{1}{m_i} \int d^3 p_i \bm{P}_i \bm{P}_i f_i(\bm{p}_i)
\end{equation}
and the heat flux vector 
\begin{equation}
    \bm{q} = \sum_{i=1}^{K} \int d^3 p_i \frac{P_i^2}{2m_i}\frac{\bm{P}_i}{m_i} f_i(\bm{p}_i).
\end{equation}
In these expressions the sum over species was explicitly written out, in future expressions
it will be implied that the sum is over all species is taken if no bounds are supplied.
This means $\sum_{i=1}^{K} \to \sum_{i}$.

\subsection{Chapman-Enskog Procedure}
\label{subsec:idea} 
The general idea of the Chapman-Enskog procedure is to expand the kinetic 
equation, Eq.~(\ref{eqn:BUU}), about equilibrium. Once this has been done, terms 
in the expansion can be ordered and the distribution function $f$ can be 
successively approximated. 

To begin, one further assumption is made, namely, that the distribution function does not 
explicitly depend on time. All time dependence arises implicitly from the hydrodynamic 
variables. This naturally leads to the definition of a $K+4$ dimensional vector,
$\vec{\beta} = (n_1, n_2, \dots, n_K, \rho v_x, \rho v_y, \rho v_y, \rho u)$,
which contains all of the hydrodynamic information. Similarly, it is assumed that 
$\vec{\Phi} = \partial \vec{\beta} /\partial t$ is not an explicit function of time either.
Thus, it is appropriate to write these quantities as 
\begin{equation}
    f_i(\bm{r}, \bm{p}_i, t) = f_i(\bm{r}, \bm{p}_i | \vec{\beta}, \nabla\vec{\beta}, \dots),
\end{equation}
and 
\begin{equation}
    \frac{\partial}{\partial t} \vec{\beta}(\bm{r}, t) = \vec{\Phi}(\bm{r}| \vec{\beta}, \nabla\vec{\beta}, \dots),
\end{equation}
where the time dependence falls in the vector $\vec{\beta}$.

Next, an expansion parameter, $\epsilon$, related to the Knudsen number of the system is 
introduced. It is supposed that 
\begin{equation}
    f_i = f_i^{(0)} + \epsilon f_i^{(1)} + \epsilon^2 f_i^{(2)} + \dots,
    \label{eqn:exp_f}
\end{equation}
and 
\begin{equation}
    \vec{\Phi} = \vec{\Phi}^{(0)} + \epsilon \vec{\Phi}^{(1)} + \epsilon^2 \vec{\Phi}^{(2)} + \dots.
    \label{eqn:exp_phi}
\end{equation}
Thermodynamics is only well defined when a system is in an equilibrium state, but 
the aim is to describe the spatial and temporal evolution of thermodynamic variables. 
To allow this, local thermodynamic equilibrium is argued for the system. This has been 
implicitly assumed since a fluid approximation was taken in the expansion about 
equilibrium. It means that all thermodynamic quantities are derived from 
the equilibrium distribution functions. In other words, perturbations of the 
distribution function contain no information on the thermodynamic state
\begin{subequations}
\begin{align}
     \int d^3 p_i f_i^{(j)} &= 0,\\
     \sum_{i}\int d^3 p_i\, \bm{p}_i f_i^{(j)} &= 0,\\
     \sum_{i}\int d^3 p_i \frac{P_i^2}{2m_i} f_i^{(j)} &= 0,
\end{align}
\label{eqn:conditions}%
\end{subequations}
for all $j > 0$. Thermodynamic fluxes do not similarly vanish. In fact,  all non-trivial contributions to the thermodynamic fluxes 
arise from higher orders of the distribution function. As such, they can be ordered
along with the distribution function
\begin{subequations}
\begin{align}
    \rho_i \bm{V}_i^{(j)} &= \int d^3 p_i \bm{P}_i f_i^{(j)},
    \label{eqn:diff_vel} \\
    \bm{\mathsf{\Pi}}^{(j)} &= \sum_{i} \bm{\mathsf{\Pi}}_i^{(j)} = \sum_{i} \frac{1}{m_i} \int d^3 p_i \bm{P}_i\bm{P}_i f_i^{(j)},
    \label{eqn:press} \\
    \bm{q}^{(j)} &= \sum_{i} \bm{q}_i^{(j)} = \sum_{i} \int d^3 p_i \frac{P_i^2}{2m_i}\bm{P}_i f_i^{(j)}.
    \label{eqn:heat_flux}
\end{align}
\label{eqn:fluxes}%
\end{subequations}
To further emphasize that the expansion is about equilibrium,
Eq.~(\ref{eqn:BUU}) is not used directly. Instead, it is reorganized to emphasize the assumed high collisionality by ordering the collision operator as a large term.
Within the realms of the expansion, this is done by adding a factor of $\epsilon^{-1}$ such that
\begin{equation}
    \frac{\partial f_i}{\partial t} 
    + \frac{\bm{p}_i}{m_i}\cdot \frac{\partial f_i}{\partial \bm{r}} 
    + \bm{F}_i\cdot \frac{\partial f_i}{\partial \bm{p}_i} = \frac{1}{\epsilon}\sum_{j}\mathcal{C}\qty(f_i,f_j).
    \label{eqn:emp_col}
\end{equation}
With this, Eqs.~(\ref{eqn:exp_f}) and (\ref{eqn:exp_phi}) are be substituted into Eq.~(\ref{eqn:emp_col})
and ordered in $\epsilon$.

\subsection{Zeroth Order Approximation}
\label{subsec:zeroth} 
The lowest order equation (order $\epsilon^{-1}$) gives
\begin{equation}
    \sum_{j}\mathcal{C}(f_i^{(0)}, f_j^{(0)}) = 0.
\end{equation}
The solution to this equation is simply when $f^{(0)}_i$ is the 
Fermi-Dirac distribution given by Eq.~(\ref{eqn:FD}) with $\delta_i = -1$~\cite{UehlingPR1933,DaligaultPOP2016,DaligaultPOP2018}. 

The fluid equations corresponding to this order of the expansion are the 
Euler equations, which are absent of any collisional transport.
At this order, solving for the thermodynamic fluxes, it is found the diffusion velocity and heat flux are identically
0, $\bm{V}^{(0)}=0$ and $\bm{q}^{(0)} = 0$. Additionally, the pressure tensor and internal energy take the values of an 
ideal Fermi gas. This implies
\begin{align}
    \rho u &= \sum_{i} \frac{3}{2}n_i \kb T \alpha_{3/2,i}, \\
    \bm{\mathsf{\Pi}} &= \mathsf{\Pi} \bm{\mathsf{I}} = \sum_{i} \mathsf{\Pi}_i \bm{\mathsf{I}} = \sum_{i} n_i \kb T \alpha_{3/2,i} \bm{\mathsf{I}}, \label{eqn:eq_press}
\end{align}
where $\alpha_{n,i} = \mathcal{Q}_{n}(\beta\mu_i)/\mathcal{Q}_{1/2}(\beta\mu_i)$ and 
$\bm{\mathsf{I}}$ is the identity tensor. The factor, $\alpha_{3/2,i}$, is the modification to the classical ideal gas due to degeneracy. It smoothly changes the average kinetic 
energy of the particles from the thermal energy, $\kb T$, in the classical 
limit, to the Fermi energy, $E_\mathrm{F} = (\hbar^2/2m_e)(3\pi^2 n_e)^{2/3}$, in the fully degenerate limit.

It is not surprising that there is no collisional transport at this order since it was assumed that
at the lowest order the system is at equilibrium. At equilibrium, the 
system is completely uniform and thermodynamic forces cannot arise to cause a thermodynamic flux. 

\subsection{First Order Approximation}
\label{subsec:first} 
The next order (order 1), is characterized by the equation
\begin{equation}
    (\mathfrak{D}f_i)^{(0)} = -\sum_{j} n_i n_j I_{ij}(\chi),
\end{equation}
where $\chi_i = f_i^{(1)}/w_i$ and $w_i = f_i^{(0)}(1-\theta_if_i^{(0)})$. Also 
$(\mathfrak{D}f_i)^{(0)}$ is the lowest order derivative in the $\epsilon$ series. Here 
the particles have been assumed to be fermions, so $\delta_i = -1$. Evaluating the derivatives of the 
equilibrium distribution function, it is obtained that
\begin{align}
    \sum_{j} n_i n_j& I_{ij}(\chi) = -w_i\Bigg[\frac{n}{n_i} \frac{\bm{P}_i}{m_i} \cdot \bm{d}_i \nonumber \\
              &+ \qty(\frac{P_i^2}{2m_i \kb T} -\frac{5}{2}\alpha_{3/2,i})\frac{\bm{P}_i}{m_i}\cdot\nabla\log T \nonumber \\
              &+\frac{1}{m_i \kb T}\qty(\bm{P}_i\bm{P}_i-\frac{1}{3}P_i^2 \bm{\mathsf{I}}):\nabla\bm{v} \Bigg]
    \label{eqn:order_1}
\end{align}
where 
\begin{equation}
    \bm{d}_i = \frac{1}{n\kb T} \qty(\nabla \mathsf{\Pi}_i - \frac{\rho_i}{\rho} \nabla \mathsf{\Pi} - \rho_i \qty(\frac{\bm{F}_i}{m_i}-\sum_{j}\frac{\rho_j}{\rho}\frac{\bm{F}_j}{m_j})),
    \label{eqn:diff_force}
\end{equation}
is the diffusion driving force, where $\mathsf{\Pi}_i$ and $\mathsf{\Pi}$ are the species and total equilibrium pressure
tensors respectively defined in Eq.~(\ref{eqn:eq_press}). There is freedom within the framework to define this 
force as above, or including the enthalpy of diffusion~\cite{Degroot, LeVanPOP2025}. 
To remain consistent with Ref.~\cite{FK}, the enthalpy of diffusion is not included into
this definition.

The integral operator $I_{ij}$ is linear and rotationally invariant.
This means $I_{ij}(F(P)\bm{P}) \propto \bm{P}$ and  allows an for 
the solution of $\chi$ to be~\cite{FK}
\begin{equation}
    \chi_i = -\frac{1}{n}\sum_{j} \bm{D}^j_i\cdot \bm{d}_j - \frac{1}{n}\bm{A}_i\cdot \nabla \log T - \frac{1}{n} \bm{\mathsf{B}} : \nabla \bm{v}
    \label{eqn:ansatz}
\end{equation}
where $\bm{D}^j_i = D^{j}\qty(P_i)\bm{P_i}$, $\bm{A}_i = A(P_i)\bm{P}_i$ and 
$\bm{\mathsf{B}} = B\qty(P_i) \qty(\bm{P}_i\bm{P}_i - 1/3 P_i^2 \bm{\mathsf{I}})$.
This form is what is needed to obtain the correct 
linear constitutive relations between thermodynamic forces and fluxes. Clearly
then, the function $D^{j}(P)$ should be related to the diffusion coefficient,
$A(P)$ should be related to the thermal conductivity, and $B(P)$ to the shear viscosity.
After inserting the solution into Eq.~(\ref{eqn:order_1}) and making correspondences 
between terms of both sides, the following relations can be found 
\begin{subequations}
\begin{align}
    &\sum_{j}\frac{n_in_j}{n^2}I_{ij}\qty(\bm{D}^k) = w_i\frac{\bm{P}_i}{n_i m_i} \qty(\delta_{ik}-\frac{\rho_i}{\rho}),
    \label{eqn:D_in_I}\\
    &\sum_{j}\frac{n_in_j}{n^2}I_{ij}\qty(\bm{A}) = \frac{w_i}{n}\frac{\bm{P}_i}{m_i} \qty(\frac{P_i^2}{2m_i\kb T} - \frac{5}{2}\alpha_{3/2,i}),\\
    &\sum_{j}\frac{n_in_j}{n^2}I_{ij}\qty(\bm{\mathsf{B}}) = \frac{w_i}{m_in\kb T}\qty(\bm{P}_i\bm{P}_i - \frac{1}{3}P_i^2\bm{\mathsf{I}}).
\end{align}
\end{subequations}
Notice that if Eq.~(\ref{eqn:D_in_I}) is multiplied by $\rho_k/\rho$ and summed 
over the index $k$, we obtain
\begin{equation}
    \sum_{k} \frac{\rho_k}{\rho} D^{k} = 0.
    \label{eqn:sum_D}%
\end{equation}
This means that the vectors $D^{k}$ are linearly dependent, which gives the transport
coefficients related to them (diffusion coefficients) summational properties.
Unfortunately this system is not closed, using the conditions (\ref{eqn:conditions}), it is 
found
\begin{subequations}
\label{eqn:sum_conditions}
\begin{align}
    \sum_{i} \int d^3 p_i w_i A(P_i)P_i^2 = 0,\\
    \sum_{i} \int d^3 p_i w_i D_i^{k}(P_i)P_i^2 = 0.
\end{align}
\end{subequations}
Now that the effect of the integral operator $I_{ij}$ on the coefficients $D^{k}$,
$A$, and $B$ has been found, they can be inserted into the bracket integral, Eq~(\ref{eqn:bracket}). Doing so shows that
\begin{subequations}
\label{eqn:all_brackets}%
\begin{align}
    &[\bm{D}^i, \bm{D}^j] = \frac{1}{\rho_i} \int d^3 p_i w_i D_i^j P_i^2,
    \label{eqn:DD_brack}\\
    &[\bm{D}^i, \bm{A}] = \frac{1}{\rho_i} \int d^3 p_i w_i A_i P_i^2,
    \label{eqn:DA_brack}\\
    &[\bm{A}, \bm{A}] = \sum_{i} \frac{1}{n m_i} \int d^3 p_i w_i A_i \qty(\frac{P_i^2}{2m_i \kb T} - \frac{5}{2} \alpha_{3/2,i}),
    \label{eqn:AA_brack}\\
    &[\bm{A}, \bm{D}^j] = \sum_{i} \frac{1}{n m_i} \int d^3 p_i w_i D_i^j \qty(\frac{P_i^2}{2m_i \kb T} - \frac{5}{2} \alpha_{3/2,i}),
    \label{eqn:AD_brack}\\
    &[\bm{\mathsf{B}}, \bm{\mathsf{B}}] = \sum_{i} \frac{1}{m_i n\kb T} \int d^3 p_i w_i B_i \frac{2}{3} P_i^4,
    \label{eqn:BB_brack}
\end{align}
\end{subequations}
where $D_i^k = D^k(P_i)$, $A_i = A(P_i)$, and $B_i = B(P_i)$, are functions of 
the magnitude of momentum. 

The tools are now in place to begin to solve for transport coefficients. Linear
constitutive relations relate thermodynamic forces to fluxes. Thermodynamic fluxes are defined
in Eq.~(\ref{eqn:fluxes}), and the forces are $\bm{d}_j$, $\nabla \log T$, and $\nabla \bm{v}$ via Eq.~(\ref{eqn:ansatz}).
Relationships between these are made by plugging the solution to the 
distribution function into the definitions of the fluxes. These result in lengthy 
integrals, which with the use of Eq.~(\ref{eqn:all_brackets}) can be written in 
terms of the bracket integrals. Beginning with the diffusion velocity,
\begin{equation}
    \bm{V}^{(1)}_i = -\sum_{j} \frac{1}{3n} [\bm{D}^i, \bm{D}^j] \bm{d}_j - \frac{1}{3n}[\bm{D}^i, \bm{A}]\nabla \log T.
\end{equation}
Transport coefficients are the coefficients relating the thermodynamic forces to
the fluxes. In this case the forces are $\bm{d}_j$ and $\nabla \log T$ and the 
flux is $\bm{V}^{(1)}_i$, so 
\begin{equation}
    D_{ij} = \frac{1}{3n} [\bm{D}^i, \bm{D}^j],
    \label{eqn:D_in_terms_brack}
\end{equation}
is the interdiffusion coefficient between species $i$ and $j$. 
Additionally, 
\begin{equation}
    D_{\textrm{T}i} = \frac{1}{3n}[\bm{D}^i,\bm{A}]
\end{equation}
is the thermal diffusion coefficient for species $i$. The
diffusion velocity is constructed by adding the zeroth and 
first order solutions. Recall though that, $\bm{V}^{(0)} = 0$,
so $\bm{V}_i = \bm{V}^{(1)}_i$ and is given by
\begin{equation}
\label{eqn:lin_diff_velo}
    \bm{V}_i = -\sum_{j} D_{ij}\bm{d}^{j} - D_{\textrm{T}i} \nabla \log T.
\end{equation}
This is the form of the linear constitutive relation which one can infer from 
nonequilibrium statistical mechanics~\cite{Degroot, LeVanPOP2025}.
Due to properties of the bracket integral and the condition Eq.~(\ref{eqn:sum_conditions}),
it can be shown that the interdiffusion coefficient is symmetric about species 
exchange, $D_{ij} = D_{ji}$, and both sets 
of diffusion coefficients are linearly dependent, which follows from Eq.~(\ref{eqn:sum_D}). This implies
\begin{equation}
    \sum_{i} \frac{\rho_i}{\rho} D_{ij} = 0,
\end{equation}
and 
\begin{equation}
    \sum_{i} \frac{\rho_i}{\rho} D_{\textrm{T}i} = 0.
\end{equation}
Further, a relationship between the interdiffusion and thermal diffusion 
coefficients can be defined as the thermal diffusion ratios, $k_{\textrm{T}i}$, where 
\begin{equation}
    \sum_{j} D_{ij} k_{\textrm{T}j} = D_{\textrm{T}i}.
    \label{eqn:diff_ratio}
\end{equation}
The thermal diffusion ratios have the condition $\sum_{i} k_{\textrm{T}i} = 0$.

Now focusing on the pressure tensor,
\begin{equation}
    \bm{\mathsf{\Pi}}^{(1)} = -\frac{\kb T}{5}[\bm{\mathsf{B}}, \bm{\mathsf{B}}]  \bm{\mathsf{S}},
\end{equation}
where $\bm{\mathsf{S}}$ is the rate of shear tensor defined as the symmetric 
traceless component of $\nabla \bm{v}$. In index notation it is given as 
$\mathsf{S}_{\alpha\beta} = (\partial_{\alpha} v_{\beta} + \partial_{\beta} v_{\alpha})/2 - (\delta_{\alpha\beta} \partial_{\gamma}v_{\gamma})/3$.
The shear viscosity is identified as 
\begin{equation}
    \eta = \frac{\kb T}{10}[\bm{\mathsf{B}}, \bm{\mathsf{B}}].
\end{equation}
Then adding the lowest order result, the total pressure tensor is
\begin{equation}
    \bm{\mathsf{\Pi}} = \mathsf{\Pi}\bm{\mathsf{I}} - 2\eta\bm{\mathsf{S}}.
\end{equation}
Finally, the heat flux is, 
\begin{align}
    \bm{q}^{(1)} = &-\frac{\kb T}{3} \sum_{j} [\bm{D}^j, \bm{A}]\bm{d}^{j} - \frac{\kb T}{3} [\bm{A},\bm{A}] \nabla \log T \nonumber \\
                   &+ \frac{5}{2}\kb T \sum_{j} n_j \alpha_{3/2,j} \bm{V}_j .
\end{align}
The partial thermal conductivity can be defined as 
\begin{equation}
    \lambda^\prime = \frac{\kb}{3} [\bm{A}, \bm{A}].
\end{equation}
Recalling $\bm{q}^{(0)} = 0$, the total heat flux is then 
\begin{equation}
    \bm{q} = - \lambda^\prime \nabla T - n\kb T \sum_{j} D_{\textrm{T}j}\bm{d}^{j} + \frac{5}{2}\kb T \sum_{j}n_j \alpha_{3/2,j} \bm{V}_i.
\end{equation}
Going one step further, the thermal diffusion ratios defined in Eq.~(\ref{eqn:diff_ratio})
can be used to define the total thermal conductivity as 
\begin{equation}
    \lambda = \lambda^\prime -n \kb \sum_{i} k_{\textrm{T}i} D_{\textrm{T}i},
    \label{eqn:total_therm}
\end{equation}
allowing the heat flux to be written in a more familiar form
\begin{equation}
    \bm{q} = -\lambda \nabla T + n\kb T \sum_{i}\qty(k_{\textrm{T}i} + \frac{5}{2}\frac{n_i\alpha_{3/2,i}}{n}) \bm{V}_i.
    \label{eqn:lin_heat}
\end{equation}

\section{Entropy Variational Principle}
\label{sec:var} 
So far, transport coefficients have been related to bracket integrals of arbitrary 
functions of momenta, $A$, $B$, and $D$. At this point, these functions are unknown
and must be approximated. 

\subsection{Entropy}
\label{subsec:entropy} 
Undergoing a similar process to linearizing the kinetic equation, the entropy for 
a quantum system can be linearized. In this manner it is found that~\cite{FK} 
\begin{align}
    \frac{\partial s}{\partial t} &= n^2 k [\chi, \chi] \nonumber \\ 
                                  &= \kb n \sum_{i,j} D_{ij} (\bm{d}_i +k_{Ti}\nabla \log T)\cdot (\bm{d}_j +k_{Tj}\nabla \log T) \nonumber \\
                                  & + \lambda \qty| \nabla \log T |^2 + \frac{2\eta}{T} \bm{\mathsf{S}}:\bm{\mathsf{S}}.
                                  \label{eqn:entrop}
\end{align}
If the system is not in equilibrium, the entropy must be increasing, thus Eq.~(\ref{eqn:entrop})
must be positive. In turn, so must the transport coefficients, $D_{ij}$, $\lambda$, and 
$\eta$. This can also be seen from their definitions in terms of bracket integrals, 
and the fact that bracket integrals are positive definite under the same argument, $[F,F] \geq 0$.
Thus, if a set of functions are introduced $\bm{d}^{k} = d^{k}\qty(P)\bm{P}$ to approximate $\bm{D}^{k}$, a good 
choice would be to require that 
\begin{equation}
    \qty[\bm{d}^{k}, \bm{d}^{l}] \leq \qty[\bm{D}^{k}, \bm{D}^{l}],
\end{equation}
so that the true rate of change of entropy can be approximated by maximizing the product $[\bm{d}^{k}, \bm{d}^{k}]$. 
To ensure this inequality, it is required that 
\begin{align}
    \sum_{j} &\frac{n_i n_j}{n^2} \qty( \qty[\bm{d}^{k}, \bm{d}^{l}]_{ij}^{\prime} +\qty[\bm{d}^{k}, \bm{d}^{l}]_{ij}^{\prime\prime}) \nonumber \\
                   &=\sum_{j} \frac{n_i n_j}{n^2} \qty( \qty[\bm{d}^{k}, \bm{D}^{l}]_{ij}^{\prime} +\qty[\bm{d}^{k}, \bm{D}^{l}]_{ij}^{\prime\prime})
                   \label{eqn:near_brack_d}
\end{align}
for all $k$ and $l$ as well as 
\begin{equation}
    \sum_{i} \frac{1}{m_i} \int d^3 p_i w_i \bm{d}_i^{k}\cdot \bm{P}_i = 0
    \label{eqn:diff_cond}
\end{equation}
for all $k$. Notice that Eq.~(\ref{eqn:near_brack_d}) is a sum over the index $i$ 
away from being a bracket integral.

Analogous trial functions, $\bm{a} = a\qty(P)\bm{P}$ and $\bm{\mathsf{b}} = b(P)\qty(\bm{P}\bm{P} - 1/3 P^2 \bm{\mathsf{I}})$ 
can be defined for the other coefficients $\bm{A}$ and $\bm{\mathsf{B}}$. The conditions 
on these trial functions are identical for $\bm{a}$. The product $\qty[\bm{a}, \bm{a}]$
is maximized such that
\begin{equation}
    \qty[\bm{a}, \bm{a}] \leq \qty[\bm{A}, \bm{A}],
\end{equation}
which requires that 
\begin{align}
    \sum_{j} &\frac{n_i n_j}{n^2} \qty( \qty[\bm{a}, \bm{a}]_{ij}^{\prime} +\qty[\bm{a}, \bm{a}]_{ij}^{\prime\prime}) \nonumber \\
                   &=\sum_{j} \frac{n_i n_j}{n^2} \qty( \qty[\bm{a}, \bm{A}]_{ij}^{\prime} +\qty[\bm{a}, \bm{A}]_{ij}^{\prime\prime})
                   \label{eqn:near_brack_a}
\end{align}
and 
\begin{equation}
    \sum_{i} \frac{1}{m_i} \int d^3 p_i w_i \bm{a}_i\cdot \bm{P}_i = 0.
\end{equation}
For $\bm{\mathsf{b}}$, again the product $\qty[\bm{\mathsf{b}}, \bm{\mathsf{b}}]$
is maximized such that,
\begin{equation}
    \qty[\bm{\mathsf{b}}, \bm{\mathsf{b}}] \leq \qty[\bm{\mathsf{B}}, \bm{\mathsf{B}}],
\end{equation}
which is guaranteed by the condition,
\begin{align}
    \sum_{j} &\frac{n_i n_j}{n^2} \qty( \qty[\bm{\mathsf{b}}, \bm{\mathsf{b}}]_{ij}^{\prime} +\qty[\bm{\mathsf{b}}, \bm{\mathsf{b}}]_{ij}^{\prime\prime}) \nonumber \\
                   &=\sum_{j} \frac{n_i n_j}{n^2} \qty( \qty[\bm{\mathsf{b}}, \bm{\mathsf{B}}]_{ij}^{\prime} +\qty[\bm{\mathsf{b}}, \bm{\mathsf{B}}]_{ij}^{\prime\prime}).
                   \label{eqn:near_brack_b}
\end{align}

\subsection{Quantum Sonine Polynomials}
\label{subsec:qsp} 
To perform the maximization procedure, the functions need to have a defined form. In 
the classical Chapman-Enskog procedure this would involve expanding the trial 
functions in terms of Sonine (Associated Laguerre) polynomials~\cite{FK}. This is a 
convenient choice because they are orthogonal with respect to Maxwellian factors, $e^{-v^2}$.
However, this is not an optimal choice here because the equilibrium distribution is Fermi-Dirac rather than Maxwellian. Instead it would be convenient to define a set of polynomials that are 
orthogonal with respect to factors $w_i = \theta_if_i^{(0)}\qty(1-\theta_if_i^{(0)})$.
Such a set of trivariate quantum polynomials was developed by Daligault~\cite{DaligaultPOP2016,DaligaultPOP2018} in his Chapman-Enskog solution of the quantum Landau Fokker-Planck equation. 

Thus, following Ref.~\cite{DaligaultPOP2018}, define a set of polynomials, $\qty{ \mathcal{L}_{\nu}^{(n)} }$
which are orthogonal with respect to the inner product 
\begin{equation}
    \langle f, g \rangle_{\nu} = \int d x\,  \omega(x) f(x) g(x),
\end{equation}
where $\omega(x)$ is defined as 
\begin{equation}
    \omega(x) = \frac{c_\Theta^{\nu+1}}{\Gamma(\nu+1)\mathcal{Q}_{\nu-1}(\beta\mu)}\frac{x^{\nu} e^{c_\Theta x - \beta\mu}}{\qty(e^{c_\Theta x - \beta\mu} + 1)^2},
\end{equation}
$c_\Theta = 1 + 1/\Theta$ and $\mathcal{Q}_{n}(\beta\mu)$ is the Fermi-Dirac
integral defined in Eq.~(\ref{eqn:fd_int}). The factor $c_\Theta$ is a
convenient scaling constant, which switches from scaling energies by 
the thermal energy, $\kb T$ in the classical limit ($\Theta \to \infty$) to the 
Fermi energy, $E_\mathrm{F}$ in the degenerate limit ($\Theta \to 0$). In Ref.~\cite{DaligaultPOP2018}, 
this factor is omitted and the quantum Sonine polynomials become numerically
difficult in the degenerate limit. Since $c_\Theta$ naturally regularizes the 
inner product in both the classical and degenerate limits, this difficulty is avoided.
It is worth noting that this inner product has been constructed such that in the 
classical limit, the inner product for the classical Sonine polynomials is recovered, 
so that the classical Chapman-Enskog result is captured.

To be further consistent with the classical Sonine polynomials, 
we take $\mathcal{L}_{\nu}^{(0)} = 1$, and the rest
of the polynomials can be constructed up to a sign with the Gram-Schmidt procedure.
A good choice of basis to orthogonalize against are the classical Sonine polynomials,
because in the classical limit, the procedure will reproduce them exactly. This means
the classical Chapman-Enskog result will be realized. Following this procedure,
one can show 
\begin{equation}
    \mathcal{L}_{\nu}^{(1)} = (\nu+1)\frac{\mathcal{Q}_{\nu}(\beta\mu)}{\mathcal{Q}_{\nu-1}(\beta\mu)} - \frac{x}{c_\Theta}.
\end{equation}
In the classical limit, when $\beta\mu \to -\infty$, the ratio $\mathcal{Q}_{\nu}(\beta\mu)/\mathcal{Q}_{\nu-1}(\beta\mu) \to 1$,
and $c_\Theta \to 1$, so that the usual Sonine polynomial is recovered.

\subsection{Trial Function Maximization}
\label{subsec:trial_d}
The quantum Sonine polynomials provide a good basis to perform a polynomial 
expansion of the product $\qty[ \bm{d}^{k}, \bm{d}^{l}]$ and maximize it. 
Suppose the value $\bm{d}^{k}$ can be expanded as a finite sum of quantum Sonine 
polynomials. In the $n$th approximation take
\begin{equation}
    \bm{d}^{k}_i = \qty(\frac{m_i}{2\kb T c_{\Theta, i}})^{1/2}\sum_{p=0}^{n-1} d_{i,p}^{k(n)} \mathcal{L}_{3/2}^{(p)}\qty(\mathcal{P}_i^2)\bm{\mathcal{P}}_i
    \label{eqn:d_expand}
\end{equation}
where $\bm{\mathcal{P}}_i = \bm{P}_i/\sqrt{2 m_i \kb T c_{\Theta, i}}$ is the normalized 
relative momentum. The quantity $\sqrt{2 m_i \kb T c_{\Theta, i}}$ is the thermal 
momentum ($p_t = \sqrt{2 m_i \kb T}$) in the classical limit, and the Fermi momentum 
($p_f = \sqrt{2 m_i E_\textrm{F}}$) in the degenerate limit. The quantity $c_{\Theta, i}$
provides a smooth transition between these two limiting cases. In this polynomial
expansion the quantities which need to be maximized are the coefficients, $d_{i,p}^{k(n)}$,
which, along with the constraint Eq.~(\ref{eqn:near_brack_d}), can be determined using 
the method of Lagrange multipliers. Ultimately, it is determined that the coefficients
$d_{l,0}^{k(n)}$ are determined by solving the system of equations 
\begin{equation}
    \sum_{j}\sum_{q=0}^{n-1} \Lambda_{ij}^{pq} d_{j,q}^{k(n)} = \frac{8}{25 \kb c_{\Theta, i}} \qty(\delta_{ik} - \frac{\rho_i}{\rho})\delta_{p0},
    \label{eqn:sys_ds}
\end{equation}
for $i = 1, \dots, K$ and $p = 0, \dots, n-1$.
The expressions $\Lambda_{ij}^{pq}$ are combinations of bracket integrals 
\begin{align}
    \Lambda_{ij}^{pq} = &\frac{8 m_i^{1/2}m_j^{1/2}}{75 \kb^2 T c_{\Theta, i} c_{\Theta, j}} \nonumber \\
                        &\Bigg\{ \delta_{ij} \sum_{h=1}^{K} \frac{n_in_h}{n^2} \qty[\mathcal{L}_{3/2}^{(q)}\qty(\mathcal{P}^2)\bm{\mathcal{P}}, \mathcal{L}_{3/2}^{(q)}\qty(\mathcal{P}^2)\bm{\mathcal{P}}]^{\prime}_{ih} \nonumber \\
                        &+ \frac{n_in_j}{n^2} \qty[\mathcal{L}_{3/2}^{(q)}\qty(\mathcal{P}^2)\bm{\mathcal{P}}, \mathcal{L}_{3/2}^{(q)}\qty(\mathcal{P}^2)\bm{\mathcal{P}}]^{\prime\prime}_{ij}\Bigg\}.
                        \label{eqn:lambda_int}
\end{align}
These ``$\Lambda$-integrals'' have a few properties that follow from the properties of 
the underlying bracket integrals. By symmetry of the bracket integral arguments
\begin{equation}
    \Lambda_{ij}^{pq} = \Lambda_{ji}^{qp}.
\end{equation}
Additionally, by conservation of momentum 
\begin{equation}
    \sum_{j} \Lambda_{ij}^{p0} = 0.
\end{equation}
These properties allow for fewer calculations in a practical implementation.
The system is supplemented by a condition that follows from Eq.~(\ref{eqn:diff_cond}),
which enforces 
\begin{equation}
    \sum_{i} \frac{\rho_i}{\rho} d_{i,0}^{k(n)} = 0.
\end{equation}
This condition prevents the system of equations (\ref{eqn:sys_ds}) from becoming 
singular.

The trial functions $\bm{a}$ are expanded in a similar way
\begin{equation}
    \bm{a}_i = \qty(\frac{m_i}{2\kb T c_{\Theta, i}})^{1/2}\sum_{p=0}^{n} a_{i,p}^{(n)} \mathcal{L}_{3/2}^{(p)}\qty(\mathcal{P}_i^2)\bm{\mathcal{P}}_i.
    \label{eqn:a_expand}
\end{equation}
Then when the product $\qty[\bm{a}, \bm{a}]$ is maximized, it results in the system
of equations 
\begin{align}
    \sum_{j} \sum_{q=0}^{n} &\Lambda_{ij}^{pq} a_{j,q}^{(n)} = \nonumber \\
                                 &\frac{4}{5\kb} \frac{n_i}{n} \frac{1}{c_{\Theta, i}^2} \qty( \frac{7}{2}\alpha_{5/2,i} - \frac{5}{2}\alpha_{3/2,i}^2)\delta_{p1}
    \label{eqn:sys_a}
\end{align}
for $i = 1, \dots, K$ and $p=0,\dots,n$, supplemented by the condition 
\begin{equation}
    \sum_{i} \frac{\rho_i}{\rho} a_{i,0}^{(n)}=0
\end{equation}
to again ensure the system is not singular.

Finally, the trial functions $\bm{\mathsf{b}}$ are expanded in
a slightly different way. Taking 
\begin{equation}
    \bm{\mathsf{b}}_i = \sum_{p=0}^{n-1} b_{i,p}^{(n)} \mathcal{L}_{5/2}^{(p)}\qty(\mathcal{P}_i^2)\qty(\bm{\mathcal{P}}_i\bm{\mathcal{P}}_i - \frac{1}{3}\mathcal{P}_i^2 \bm{\mathsf{I}})
\end{equation}
and minimizing the bracket integral $\qty[\bm{\mathsf{b}},\bm{\mathsf{b}}]$, gives 
the system of equations
\begin{equation}
    \sum_{j}\sum_{q=0}^{n-1} H_{ij}^{pq}b_{j,p}^{(n)} = \frac{2}{\kb T} \frac{n_i}{n}\frac{\alpha_{3/2,i}}{c_{\Theta, i}}\delta_{p0}
    \label{eqn:h_b}
\end{equation}
where the coefficients $H_{ij}^{pq}$ are defined as 
\begin{align}
    H_{ij}^{qr} &= \nonumber \\
                &\frac{2}{5\kb T} \Bigg\{ \delta_{ij}\sum_{h=0}^{K} \frac{n_in_h}{n^2} \qty[\mathcal{L}_{5/2}^{(q)}\qty(\mathcal{P}^2) \bm{\mathsf{P}},\mathcal{L}_{5/2}^{(r)}\qty(\mathcal{P}^2) \bm{\mathsf{P}}]_{ih}^\prime \nonumber \\
                & + \frac{n_in_j}{n^2}[\mathcal{L}_{5/2}^{(q)}\qty(\mathcal{P}^2) \bm{\mathsf{P}},\mathcal{L}_{5/2}^{(r)}\qty(\mathcal{P}^2) \bm{\mathsf{P}}]_{ij}^{\prime\prime} \Bigg\},
\end{align}
and the tensor $\bm{\mathsf{P}} = \bm{\mathcal{P}}\bm{\mathcal{P}} - \mathcal{P}^2 \bm{\mathsf{I}}/3$. 

At this point, each of the bracket integrals is expressed as a sum of quantum polynomials. 
The final step is to connect these back to the transport coefficients. 

\subsection{Transport Coefficients}
\label{subsec:transport}
Using the same expansions as in Section~\ref{subsec:trial_d}, the transport coefficients
can be written in terms of the expansion coefficients. The process will be explained for 
the diffusion coefficient, as the rest follow the same procedure. In the maximization,
the product $\qty[\bm{d}^{k}, \bm{d}^{l}]$ was used to approximate $\qty[ \bm{D}^{k}, \bm{D}^{l}]$.
This was done under the condition of Eq.~(\ref{eqn:near_brack_d}), which if a sum 
over the index $i$ is taken, becomes
\begin{equation}
    \qty[\bm{d}^{k}, \bm{d}^{l}] = \qty[\bm{d}^{k}, \bm{D}^{l}].
\end{equation}
This is helpful, as in this form the expansion of $\bm{d}^{k}$, Eq.~(\ref{eqn:d_expand}),
can be used in Eq.~(\ref{eqn:DD_brack}) to find that 
\begin{equation}
    \qty[\bm{d}^{k}, \bm{d}^{l}] = \frac{3}{2} \frac{1}{c_{\Theta, l}} d_{l,0}^{k(n)} = \frac{3}{2} \frac{1}{c_{\Theta, k}} d_{k,0}^{l(n)},
\end{equation}
where the second equality comes from symmetry of the arguments of the bracket integral.
The diffusion coefficient in the $n$th non-zero approximation can be determined 
using Eq.~(\ref{eqn:D_in_terms_brack}) to be
\begin{equation}
    \qty[D_{ij}]_{n} = \frac{1}{2nc_{\Theta, i}} d_{i,0}^{j(n)} = \frac{1}{2n c_{\Theta, j}} d_{j,0}^{i(n)}.
\end{equation}
In this expression, $d_{i,0}^{j(n)}$ is determined by solving the system of 
equations (\ref{eqn:sys_ds}) and thus solving the bracket integrals of the 
form of Eq.~(\ref{eqn:lambda_int}). 

The same procedure is repeated for the other transport coefficients. Doing so, the thermal diffusion coefficient is
\begin{align}
\label{eq:DT}
    [D_{\textrm{T}i}]_n &= -\frac{1}{2nc_{\Theta, i}} a_{i,0}^{n} \nonumber \\
                        &= -\frac{5}{4n} \sum_{i} \frac{n_i}{nc_{\Theta, i}^2}d_{i, 1}^{k(n+1)} 
                         \qty( \frac{7}{2}\alpha_{5/2,i} - \frac{5}{2}\alpha_{3/2,i}^2).
\end{align}
The first expression uses Eq.~(\ref{eqn:DA_brack}) and expands the trial function $\bm{a}$,
whereas the second expression uses Eq.~(\ref{eqn:AD_brack}) and expands the trial 
function $\bm{d}$. These are equivalent by the symmetry of the bracket integral 
arguments. 

Note that in the second equality in Eq.~(\ref{eq:DT}) the expansion is taken to order $n+1$.
This is done to be consistent in the ordering of the transport coefficients with 
Ref.~\cite{FK}. There, order $n$ corresponds to the first non-zero order 
of the polynomial expansion, not the polynomial degree of the expansion. 
This means $\qty[D_{ij}]_1$ has a contribution due to degree 0 polynomials, whereas 
$\qty[D_{\textrm{T}i}]_1$ does not and the first non zero term comes from degree 1 polynomials.
This is the reason the expansion of $\bm{d}$ is carried out to degree $n-1$ in Eq.~(\ref{eqn:d_expand}), but 
the expansion of $\bm{a}$ is done to degree $n$ in Eq.~(\ref{eqn:a_expand}). Overall, due to this slight inconsistency,
if transport coefficients are to be combined, they must be either carried out to 
the same polynomial order, or close to convergence in the polynomial expansion. 
For example, the total thermal conductivity to order $n$ must use diffusion coefficients
of order $n+1$, but thermal diffusion coefficients of order $n$ to perform the expansion to a consistent 
polynomial degree. 

This is mentioned in  Ref.~\cite{FK} when the thermal diffusion coefficient is referred to as a ``second-order 
transport effect". Physically, the interpretation is that interdiffusion 
processes are captured by momentum transfer phenomena to lowest order. In fact, this is what 
the lowest order of the polynomial expansion is. Plugging in a degree 0 quantum 
Sonine polynomial to the bracket integral results in an expression like 
$\bm{\mathcal{P}} - \bm{\mathcal{P}}^\prime$ in the integrand. This is a simple
momentum transfer, and it thus is not expected to contribute to energy transfer 
processes like thermal diffusion, or thermal conductivity. In this way, one may 
interpret the odd combination of orders needed to combine transport coefficients
as simply matching the correct physics effects. 

Alternatively, this can be viewed in terms of the distribution function. So far,
the perturbation to the distribution function has been solved for and expanded in 
a polynomial basis. Therefore, to be consistent with the applied perturbation, the 
polynomial used must be the same. This means interdiffusion coefficients must be 
carried out to the $n+1$ order if the rest of the coefficients are carried out to 
the $n$th order.

The expressions for the partial thermal conductivity and shear viscosity follow 
as an application to the explained procedure to Eq.~(\ref{eqn:AA_brack}) and 
Eq.~(\ref{eqn:BB_brack}) respectively. For the partial thermal conductivity it is 
found that 
\begin{equation}
    \qty[\lambda^\prime]_n = \frac{5\kb}{4}\sum_{i}\frac{n_i}{n c_{\Theta, i}^2} a_{i,1}^{(n)} \qty( \frac{7}{2}\alpha_{5/2,i} - \frac{5}{2}\alpha_{3/2,i}^2),
\end{equation}
and the shear viscosity is given by
\begin{equation}
    \qty[\eta]_n = \frac{\kb T}{2} \sum_{i} \frac{n_i}{n}\frac{\alpha_{3/2,i}}{c_{\Theta, i}} b_{i,0}^{(n)}.
\end{equation}
Recall, the total thermal conductivity can be constructed from the partial thermal conductivity,
the thermal diffusion coefficient, and the interdiffusion coefficient with Eq.~(\ref{eqn:total_therm}).
The dependence on the interdiffusion coefficient arises in the definition of the 
thermal diffusion ratios.

\section{Electronic Transport Coefficients in an Ion-Electron Plasma}
\label{sec:electrons}
The kinetic equation and resulting fluid description derived in the previous sections is a general result and 
in theory could be applied to a variety of fermionic gases with arbitrary potentials 
and with any number of present species. In this work, the system of interest is a 
plasma of $N$ ion species with free electrons. Specifically, within this system,
the particular interest is in calculating quantities which involve the electrons. In warm dense 
matter the electrons must be treated quantum mechanically, which is why the use of the 
BUU kinetic equation is necessary. These transport properties include 
electrical conductivity, thermal conductivity, and the electrothermal
coefficient. Other transport quantities, the ionic diffusion and 
shear viscosity, are dominated by the classical ionic species for which the theory reduces 
to classical mean force kinetic theory~\cite{BaalrudPRL2013,BaalrudPOP2019}. Excellent agreement for classical mean force 
kinetic theory in the warm dense matter regime has been shown previously~\cite{DaligaultPRL2016, BerrensPRE2026}
and will not be revisited here.

From this point forward, a mixture of $N$ ionic species (with index $i$) and 
electrons (with index $e$) will be considered. For an ion-electron plasma, a more 
valuable quantity than the diffusion velocity, $\bm{V}$ is the current density
$\bm{J}$. The linear constitutive relation for $\bm{J}$ can be found using 
that for $\bm{V}$ in Eq.~(\ref{eqn:lin_diff_velo})
\begin{align}
    \bm{J} =& n_e q_e \bm{V}_e + \sum_{i=1}^{N} n_i q_i \bm{V}_i
\end{align}
where $q_i$ is the charge of species $i$. Recall, in Eq.~(\ref{eqn:lin_diff_velo}) for the 
diffusion velocity, the diffusion driving force, $\bm{d}^j$ defined in Eq.~(\ref{eqn:diff_force})
must be known. In turn, this diffusion driving force depends on forces not explicitly included
the the thermodynamic state of the system. Generally, this could include effects like gravity, 
but in the case of an unmagnetized plasma, the dominant term is the electrostatic force from 
the local electric field in the plasma, $\bm{E}$. If the sums are expanded and the terms small in the mass ratio between the electrons and ions are dropped ($m_e/m_i \ll 1$),
then the result is 
\begin{equation}
    \bm{J} = \sigma \qty(\bm{E} - \frac{\nabla \mathsf{\Pi}_e}{n_e q_e}) - \varphi \nabla T
\end{equation}
where $\mathsf{\Pi}_e = n_e \kb T \alpha_{3/2,e}$ is the partial electronic pressure
and $\sigma$ is the electrical conductivity defined as 
\begin{equation}
    \sigma = \frac{q_e^2 n_e^2}{n \kb T} D_{ee}.
\end{equation}
Note that by the linear dependence of interdiffusion coefficients 
\begin{equation}
    D_{ee} = -\sum_{i=1}^N \frac{\rho_i}{\rho_e} D_{ei}
\end{equation}
where the sum is over ionic species. This means that $D_{ee}$ actually contains 
all the information of how electrons diffuse through all other species and themselves.

The electrothermal coefficient, $\varphi$, is defined as 
\begin{equation}
    \varphi = \frac{n_e q_e D_{\textrm{T}e}}{T}.
\end{equation}
The mass ratio expansion is taken in this situation because the smallest ratio between
an electron and ion would be that of a hydrogen plasma (protons and electrons), 
where the mass ratio is $m_e/m_i \approx 5.4\times 10^{-4} \ll 1$. Any other 
physical system will have more massive ions and only make this ratio smaller, so
any time this expansion can be exploited, it will be used. 

Using Cramer's rule to solve the linear system Eq.~(\ref{eqn:sys_ds}), the electrical
conductivity can be written explicitly in terms of the $\Lambda$ integrals, Eq.~(\ref{eqn:lambda_int}).
In the third approximation it is given by
\begin{equation}
    \qty[\sigma]_{3} =  \frac{4q_e^2 n_e^2}{25 n^2 \kb Tc_{\Theta, e}^2} \frac{\Lambda^{11}_{ee}\Lambda^{22}_{ee} - \qty(\Lambda^{12}_{ee})^2}{\det(\bm{\mathsf{M}})},
    \label{eqn:elec_conduct}
\end{equation}
where $\bm{\mathsf{M}}$ is a matrix of the $\Lambda$ integrals defined as
\begin{equation}
    \bm{\mathsf{M}} = 
    \begin{pmatrix}
        \Lambda^{00}_{ee} & \Lambda^{01}_{ee} & \Lambda^{02}_{ee} \\
        \Lambda^{10}_{ee} & \Lambda^{11}_{ee} & \Lambda^{12}_{ee} \\
        \Lambda^{20}_{ee} & \Lambda^{21}_{ee} & \Lambda^{22}_{ee} 
    \end{pmatrix}.
\end{equation}
This expression was obtained after a mass ratio expansion of the linear system 
Eq.~(\ref{eqn:sys_ds}). It may seem that this expression is only on the electrons in 
the system, however this is not true. In the definition of the 
$\Lambda$ integrals there are three types of bracket integrals. The first ones are
of the form $\qty[ \cdot, \cdot]_{ei}^{\prime}$ which contain the electron-ion 
interaction for each ionic species. That leaves two integrals, one of the form
$\qty[ \cdot, \cdot]_{ee}^{\prime}$ and the other $\qty[\cdot, \cdot]_{ee}^{\prime\prime}$
Together they can be cast into a single species bracket integral containing only 
electron-electron interactions. This means overall the expressions, $\Lambda^{pq}_{ee}$ 
examine how the electron's momenta and energy change due to collisions which 
can be due to collisions with ions ($\qty[ \cdot, \cdot]_{ei}^{\prime}$) or other 
electrons (the other two bracket integrals).  

The other species bracket integrals drop out the expression because if ions are 
treated as much more massive than electrons,  
the ion's velocity changes little due to a single interaction with an electron. The integrals 
$\Lambda_{ei}^{pq}$ and $\Lambda_{ie}^{pq}$ represent how the ion's velocity distribution 
changes as a result of these interactions, and thus are higher order corrections. The quantities 
$\Lambda_{ij}^{pq}$ for pairs of ionic species $i$ and $j$, represent ionic momentum and energy 
exchange among themselves. Due to the mass of the ions though, they move slower than electrons 
on average and interact among themselves on a much longer time scale than electrons do. Thus,
velocity (or current) dependent transport will be dominated by the electronic motion.
In the mass ratio expansion, this manifests in the form of the 
$\Lambda_{ij}^{pq}$ quantities completely canceling out of the expression, leaving only
$\Lambda_{ee}^{pq}$ terms.

Using the same arguments, but solving the system of equations, Eq.~(\ref{eqn:sys_a}),
the electrothermal and thermal conductivity coefficients can be written in the 
second approximation. Recall, the second approximation of these transport coefficients
corresponds to the same polynomial order as the third approximation to the electrical 
conductivity. The electrothermal coefficient is 
\begin{align}
    \qty[\varphi]_2 = \frac{2 q_e n_e^2}{5 \kb Tn^2 c_{\Theta, e}^3} &\qty(\frac{7}{2} \alpha_{5/2,e} - \frac{5}{2}\alpha_{3/2,e}) \nonumber \\
    \times&\frac{\Lambda_{ee}^{01} \Lambda_{ee}^{22} - \Lambda_{ee}^{02}\Lambda_{ee}^{21}}{\det(\bm{\mathsf{M}})}
\end{align}
and the thermal conductivity is 
\begin{align}
    \qty[\lambda]_2 = \qty(\frac{n_e}{n c_{\Theta, e}^2})^2&\qty(\frac{7}{2} \alpha_{5/2,e} - \frac{5}{2}\alpha_{3/2,e})^2 \nonumber \\
    \times&\frac{\Lambda_{ee}^{22}}{\Lambda_{ee}^{11}\Lambda_{ee}^{22} - \qty(\Lambda_{ee}^{12})^2}.
\end{align}
In all of these expressions, the $\Lambda$ integrals are computed from Eq.~(\ref{eqn:lambda_int}),
as sums of bracket integrals. Reduced forms of the bracket integrals can be found 
in Appendix~\ref{sec:brackets}.

\section{Results and Discussion}
\label{sec:results} 

Highlights of the results of this model in comparison to DFT-MD and experimental data is shown in the companion paper~\cite{BabatiPRL2026} for hydrogen and aluminum. 
Here, a deeper investigation into the hydrogen data is provided, along with additional calculations for carbon and beryllium. 

\subsection{Hydrogen}
\label{subsec:h} 

\begin{figure}
    \includegraphics[width=0.48\textwidth]{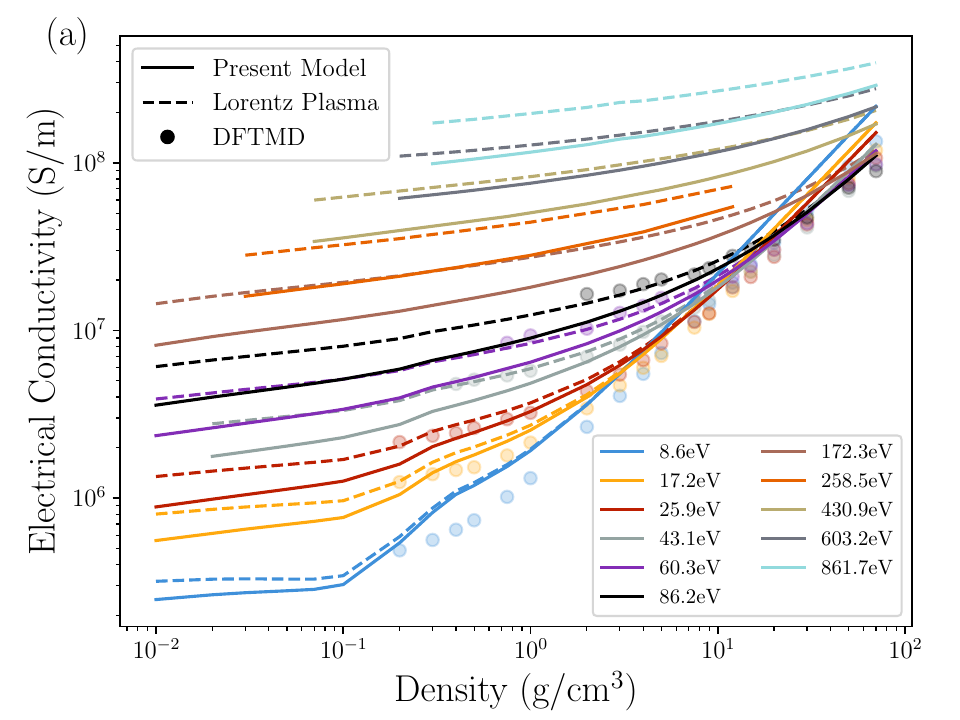}
    \includegraphics[width=0.48\textwidth]{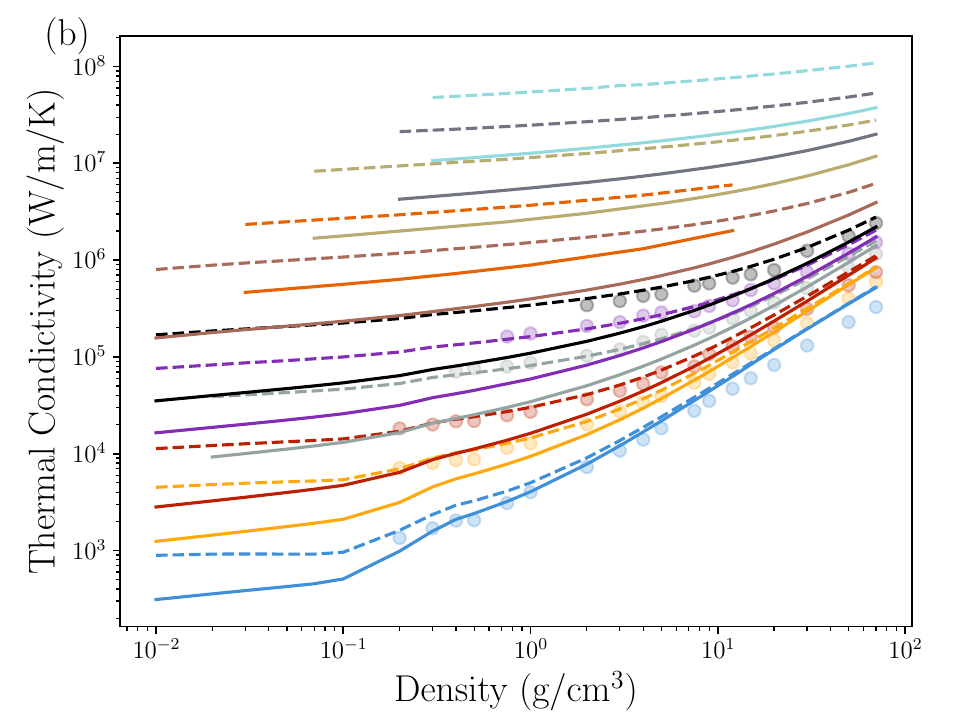}
    \includegraphics[width=0.48\textwidth]{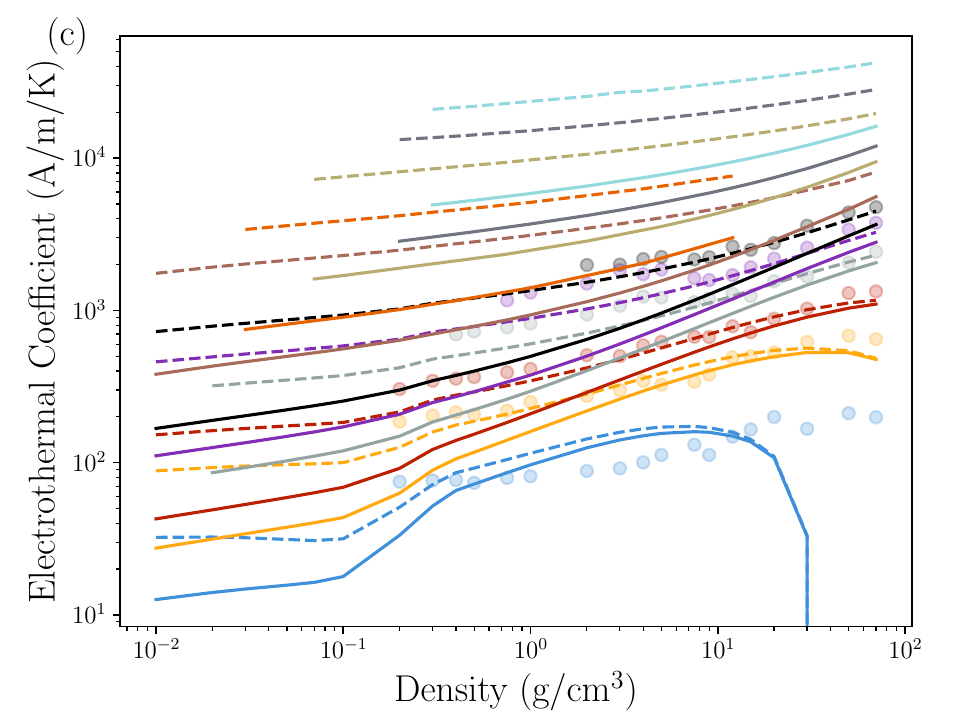}
    \caption{\label{fig:hydrogen_data} Electronic transport coefficients plotted 
    as a function of mass density for various isotherms of hydrogen ranging from 8.6~eV to 
    860~eV. The panels are (a) electrical conductivity, (b) thermal conductivity,
    and (c) the electrothermal coefficient. The solid lines are predictions of the 
    present model, dotted lines are predictions of the present model with electron-electron interaction terms turned off (i.e., the Lorentz plasma), and the points 
    are obtained from  DFT-MD simulations from Ref.~\cite{BergermannPOP2026}. Data from the DFT-MD simulations is restricted to the lower half of 
    temperatures, from 8.6~eV to 86~eV.}
\end{figure}

Much focus has been placed on the calculation of transport coefficients of hydrogen~\cite{GrabowskiHEDP2020, StanekPOP2024,BergermannPOP2026}.
Not only is it crucial to inertial confinement fusion~\cite{NIF, GomezPRL2014},
but also to astrophysical systems like dense stars~\cite{Chabrier_Schatzman_1994, SaumonPhysRep2022}.
Hydrogen is also atomically simple, removing much of the complicated atomic physics
from the calculation of transport properties in many cases.
Additionally, a dataset of transport coefficients ranging from the condensed matter
to the cold side of warm dense matter was produced using DFT-MD~\cite{BergermannPOP2026},
making comparisons to the present model convenient. To make the comparison,
all three electronic transport coefficients were computed from densities ranging 
from 0.01~g/cm$^{3}$ to 70~g/cm$^{3}$ and temperatures ranging from 8.6~eV to 860~eV. These 
conditions roughly correspond to $\Gamma = 0.01-10$ and $\Theta=100-0.02$. The DFT-MD
data is limited to the lower temperatures and higher densities, reaching a maximum 
temperature of 86~eV and a minimum density of 0.2~g/cm$^{3}$. 

This comparison is shown in Fig.~\ref{fig:hydrogen_data}. Overall, good qualitative agreement
can be seen, particularly at lower temperatures and higher densities. In this regime, DFT-MD 
is expected to perform very well, and the present theory agrees well. Since the present model is
based on plasma theory, it can access higher temperatures and lower densities than DFT-MD.
This is why the comparison is limited to temperatures less than 100~eV and densities greater 
than 0.2~g/cm$^3$. Small differences between the present theory and the DFT-MD are present, but 
this is most likely due to non-ideal free electrons, and that molecular hydrogen could be present at the most degenerate 
conditions. In the regions where the system is more 
plasma-like (mostly ionized) the theory and DFT-MD agree well.


At higher temperatures, the DFT-MD begins to deviate from the full model predictions and instead follows the ``Lorentz Plasma'' model. 
This is an evaluation of the present theory that turns off electron-electron collisions. 
Similar Lorentz plasma models that treat only the electron-ion interaction contributions to the conductivity coefficients are popular in the condensed matter limit because Pauli blocking 
limits interactions between electrons~\cite{Ashcroft, BabatiPRE2026}. In plasma 
physics, this is not the correct description, as it is known that electron-electron interactions
can change the transport properties by a factor of two or more~\cite{SpitzerPhysRev1953}.
The electron-electron interaction contributions are commonly known as the ``Spitzer correction". With the present
theory it is simple to toggle between the Lorentz plasma and the fully interacting 
system by turning the electron-electron collision terms off. In Eq.~(\ref{eqn:lambda_int}), if the terms involving only electron-electron collisions are dropped the Lorentz limit is reached, leaving
\begin{align}
    \Lambda_{22, L}^{pq} = \sum_{i=1}^N\frac{8 m_e n_in_e}{75n^2 \kb^2 T c_{\Theta, e}^2} \qty[\mathcal{L}_{3/2}^{(p)}\qty(\mathcal{P}^2)\bm{\mathcal{P}}, \mathcal{L}_{3/2}^{(q)}\qty(\mathcal{P}^2)\bm{\mathcal{P}}]^{\prime}_{ei}
                        \label{eqn:lorentz_lambda_int}
\end{align}
where the sum is over ionic species and the expressions for the transport coefficients themselves remain unchanged.

The good agreement between the Lorentz model evaluation and DFT-MD suggests that DFT-MD does not include electron-electron interaction contributions. 
This is not surprising, and has been explained in previous investigations~\cite{FrenchPRE2022, BergermannPOP2026}. There are two main 
reasons for this. The first is that density functional theory (DFT) is a mean field theory.
In DFT, the many-body quantum system is mapped to a system of interaction-free 
electrons in an effective potential. This system does not contain explicit electron-electron 
interactions, but treats them through the mean field. The exchange-correlation potential is 
intended to include fill this gap and re-include electron-electron interactions, but must 
be determined from outside the framework of DFT and is in general impossible to know for 
all systems.
In
degenerate conditions, $\Theta < 0.1$, electronic interactions are small and exchange 
correlation potentials are relatively simple to approximate~\cite{PerdewAIPCP2001, SunPRL2015}. 
In this case, DFT overall becomes a very good approximation. On the other hand, outside of this regime, 
exchange correlation potentials become extremely complicated and computationally
prohibitive to calculate~\cite{MoldabekovJCTC2024}. There have been recent developments to 
add these interactions back to DFT~\cite{RobinsonPRL2026}, but so far these have not demonstrated that they can extend far into the warm dense matter regime.

The second reason DFT-MD corresponds to a Lorentz plasma involves the Kubo-Greenwood
formulation~\cite{Kubo, Greenwood}. This procedure is ultimately how conductivity coefficients 
are calculated from electronic wavefunctions in DFT. One defining feature of the Kubo-Greenwood 
formula is that it makes the same non-interacting approximation that DFT does~\cite{Kubo, Greenwood}. 
This approximation is fitting for typical DFT in the condensed matter regime, which makes the 
same assumption, but ultimately limits the applications to systems where electron-electron 
interactions are negligible.

\begin{figure}
    \includegraphics[width=0.48\textwidth]{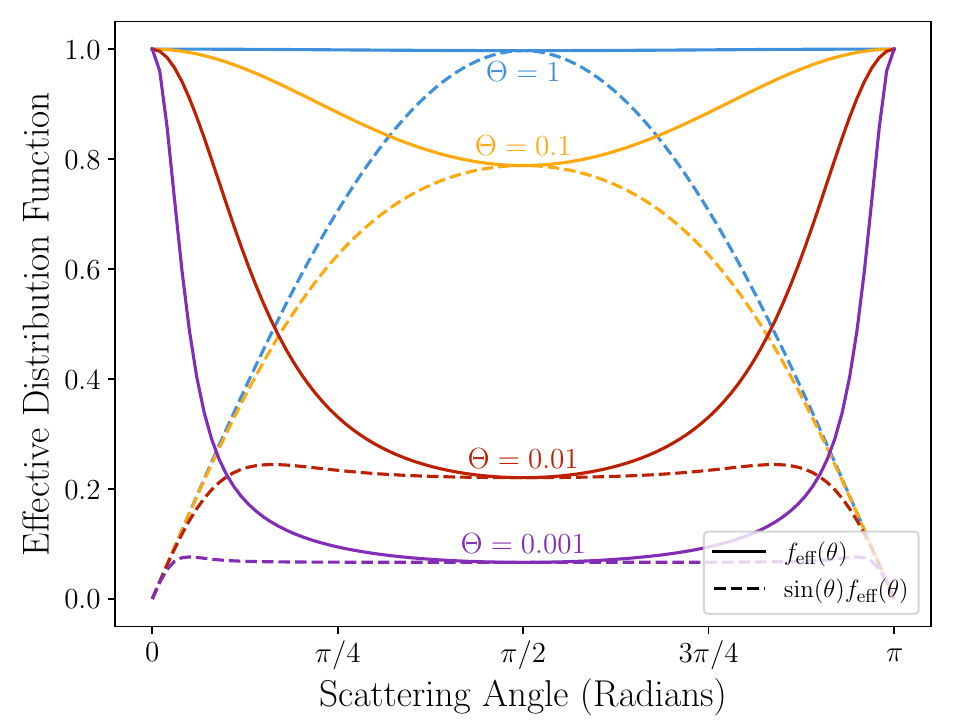}
    \caption{\label{fig:dist} The effective distribution function for an electron-electron 
    interaction plotted as a function of scattering angle, $\theta$. 
    This shows the relative likelihood of a collision occurring at a given scattering angle. The dashed 
    lines multiply this effective distribution by a weight of $\sin\theta$, which is 
    the fraction of solid angle each scattering angle occupies. When the system is classical,
    there is no preference for scattering angle, but at large degeneracy, collisions are 
    limited to forward or backscatter, which when weighted by the $\sin$ function means 
    the collision does not occur.} 
\end{figure}

The reason that electron-electron interactions become negligible in the degenerate regime 
is due to the Pauli exclusion principle. This is illustrated in Fig.~\ref{fig:dist}, where the effective 
distribution function, $f_{\textrm{eff}}$, for degenerate electrons is plotted as a function of scattering
angle and normalized so its maximum is 1. This quantity represents the statistical contribution to the relative probability that an electron-electron
scattering event occurs with a specific scattering angle, $\theta$. The collision frequency for any of the transport coefficients is  
determined from this quantity by multiplying by the differential scattering cross section and then 
integrating over scattering angle and a function of the relative energy of the collision. It is 
defined by Eq.~(\ref{eqn:effective_dist}) by setting $F_{pq} = 1$. When $\Theta \ll 1$,
scattering events are limited to only forward or backscattering as shown by the peaks 
around scattering angles of $0$ and $\pi$. This is ultimately a geometric constraint  requiring the length of the electron's momentum vectors to be the Fermi momentum.
Then, when the effective distribution function is angularly integrated, it is weighted 
by $\sin\theta$, which is the fraction of solid angle for a given scattering angle.
Overall, the contribution of electron-electron interactions to the transport of momentum and energy 
becomes negligible in the degenerate limit. When $\Theta \geq 1$,
there is no dependence on scattering angle because the length of the electron momentum 
vectors are not restricted by Pauli blocking, meaning they can contribute to energy transfer. 
In classical statistical regimes, this has the effect of reshaping the electron distribution 
function, which then influences the transport coefficients.

\begin{figure}
    \includegraphics[width=0.48\textwidth]{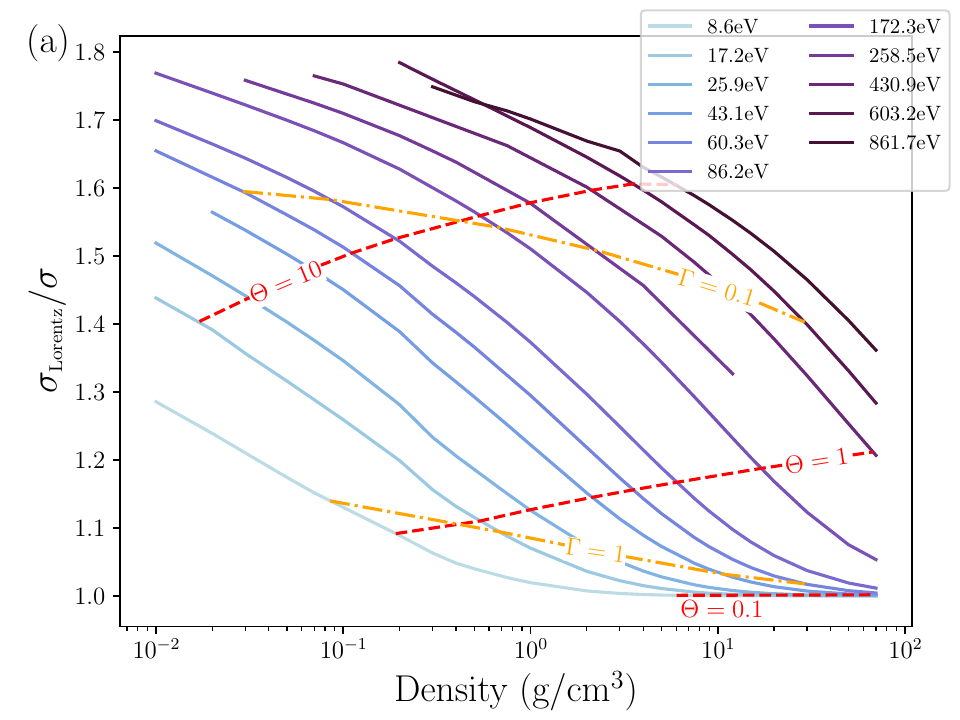}
    \includegraphics[width=0.48\textwidth]{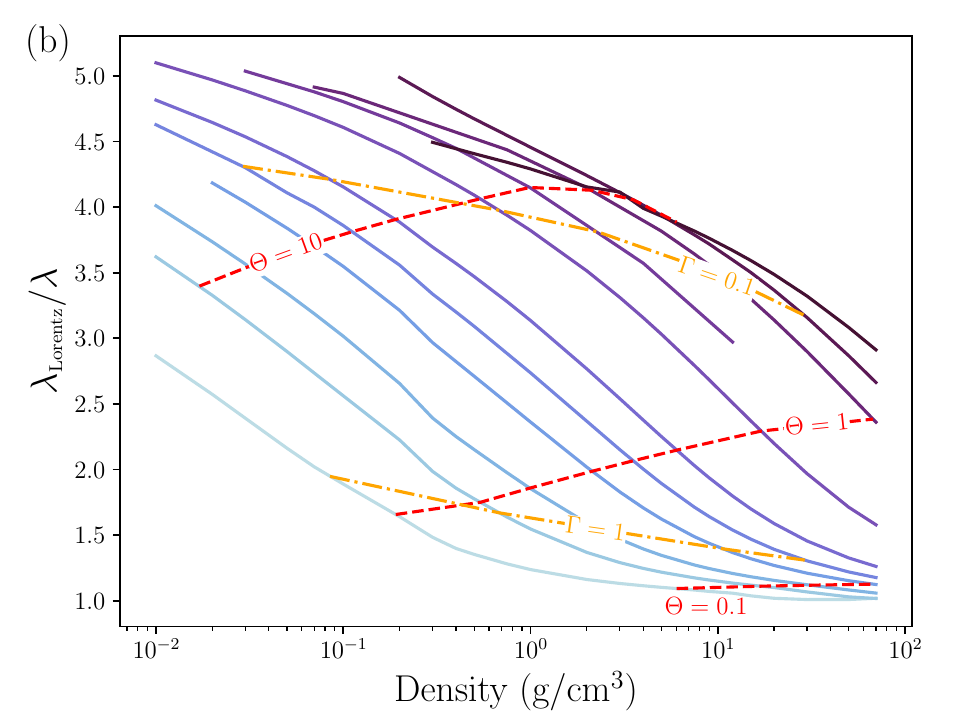}
    \includegraphics[width=0.48\textwidth]{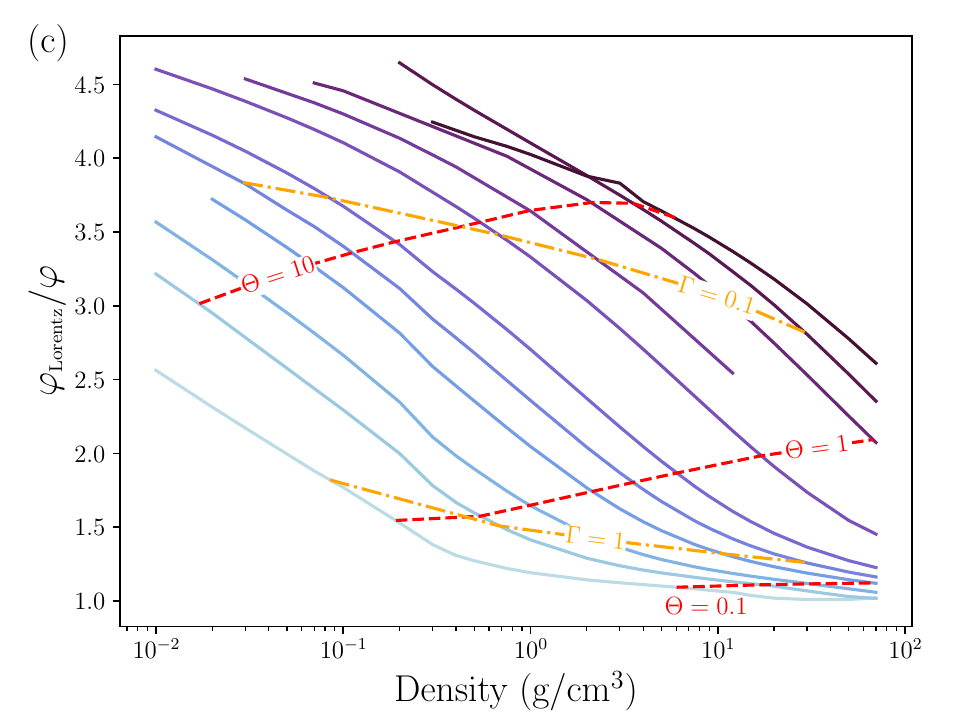}
    \caption{\label{fig:ratio}Ratio of the ``Lorentz plasma'' (electron-ion interaction only) and full model evaluations  for the (a) electrical conductivity,
    (b) thermal conductivity, and (c) electrothermal coefficient. This is the same data as in Fig.~\ref{fig:hydrogen_data}. Predictions are plotted as a function of 
    mass density for various temperatures ranging from 8.6~eV to 860~eV. Red dashed lines
    denote where $\Theta$ takes the values of 0.1, 1, and 10 for each isotherm. Similarly, gold dashed dotted lines show where $\Gamma_e=0.1$ and 1.  The figure shows that the full model results asymptote to the Lorentz plasma limit when $\Theta \lesssim 0.1$.}
\end{figure}

In warm dense matter, the contribution of electron-electron collisions to electronic transport coefficients can be large. This is demonstrated in Fig.~\ref{fig:ratio}, which shows the ratio between the Lorentz plasma model
and full model. 
Focusing first on warm dense matter parameters, this ratio is around a factor of two when $\Theta \approx 1$ for the thermal conductivity and the electrothermal coefficients. 
It is around 1.2 for electrical conductivity at the same $\Theta$ value. 
As expected, the ratio for every coefficient gets largest in the classical weakly coupled limit ($\Theta \gg 1$, $\Gamma \ll 1$), and trends toward unity in the strongly coupled degenerate limit ($\Gamma \gg 1$, $\Theta \ll 1$).  
The comparative magnitude of electron-electron contributions differs between coefficients because 
the physical processes that determine electrical conduction are different from thermal conduction and 
the electrothermal effect. Specifically, electrical conductivity depends on momentum exchange phenomena, of which 
electron-electron interactions do  not contribute to lowest order.  Alternatively, thermal conductivity and 
electrothermal processes are dependent on energy exchange to first order, which does have an 
electron-electron component. Thus, when the Lorentz limit of the electrical conductivity is taken
it only affects higher order corrections to the value, whereas it affects 
the lowest order value of thermal conductivity and the electrothermal coefficient.
Due to this, DTF-MD transport coefficients could be up to a factor of 2-5 off from the correct value, 
particularly at high temperature. Overall, this could lead to a significant underestimation of thermal 
diffusion timescales in hydrodynamic simulations, yielding misleading results when designing experiments.

\subsection{Carbon}
\label{subsec:c} 
\begin{figure}
    \includegraphics[width=0.48\textwidth]{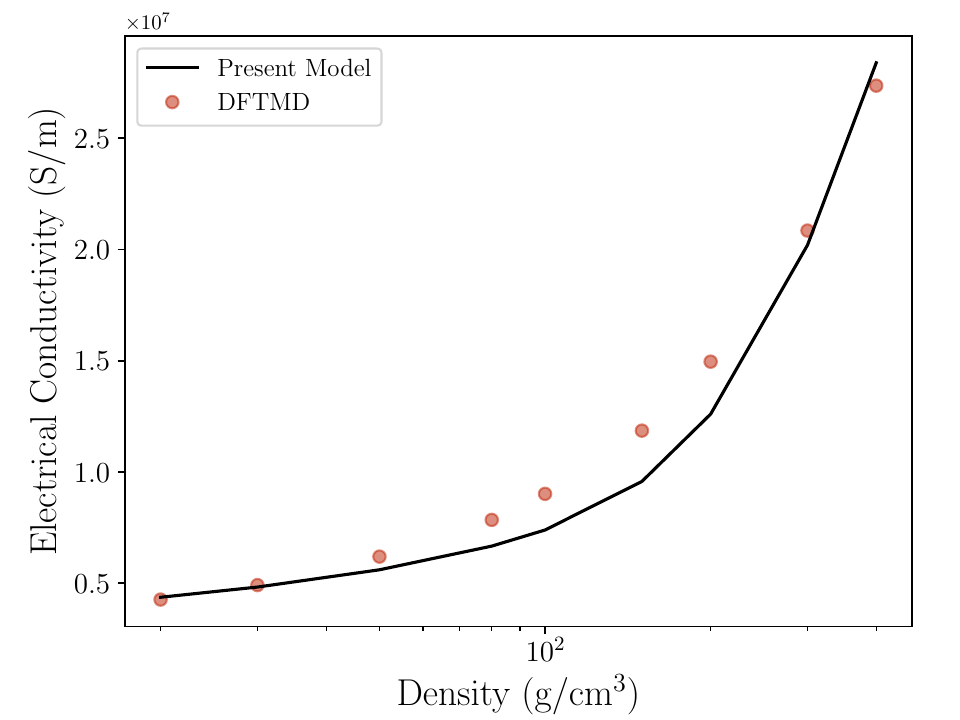}
    \caption{\label{fig:carbon_elec} Electrical conductivity of Carbon at 100~eV calculated
    with Eq.~(\ref{eqn:elec_conduct}) as a function of mass density. These are compared to DFT-MD simulations
    from Ref.~\cite{BethkenhagenPRR2020} where electrical conductivity was calculated as
    an intermediate step. The conditions are all in the degenerate regime so
    electron-electron interactions are negligible.}
\end{figure}

Current capsule designs for fusion experiments on the NIF use an ablator made of 
high density carbon~\cite{NIF}. During an implosion, this ablator transitions from 
a solid, through warm dense matter, into a plasma. Understanding the transport properties 
of carbon is important to predicting the experiments beforehand, and in interpreting 
the data afterwards. In light of this, there have been 
DFT-MD simulations~\cite{BethkenhagenPRR2020} which calculated the electrical 
conductivity of carbon as an intermediate step in the study. Figure~(\ref{fig:carbon_elec})
shows a comparison between the DFT-MD data and the present model for carbon at 100~eV
and from 20~g/cm$^{3}$ to 300~g/cm$^{3}$. 

For all data, $\Gamma \approx 1$ to 10 and $\Theta \approx 1$ to 0.1. This places
the conditions calculated within the warm dense matter regime. Since $\Theta \lesssim 1$
for all the conditions, it is expected that electron-electron interactions are relatively 
small. This means that the DFT-MD provides a good benchmark to test the model against. 

It can be seen in Fig.~\ref{fig:carbon_elec} that the agreement is good across the densities 
calculated, as the numerical values are all within an order unity.
As the density increases, differences between the model and the DFT-MD begin to 
grow. At extreme densities, carbon can display complicated electronic structure, 
which the average atom model cannot capture. Despite this possibility, the differences are not 
very significant, indicating that the present model can capture the physics relevant for 
carbon at these conditions. It should be noted, however, the  DFT conductivity values shown 
here were derived from the simulation dataset of Bethkenhagen et al.~\cite{BethkenhagenPRR2020}, which was 
primarily generated for ionization degree calculations. A limited particle-number 
convergence analysis was performed and used to guide a Drude-model extrapolation. Nevertheless, 
since the calculations were not specifically designed as a dedicated DC conductivity study, some 
uncertainty in the absolute conductivity values may remain.

\subsection{Beryllium}
\label{subsec:be}

\begin{figure}
    \includegraphics[width=0.48\textwidth]{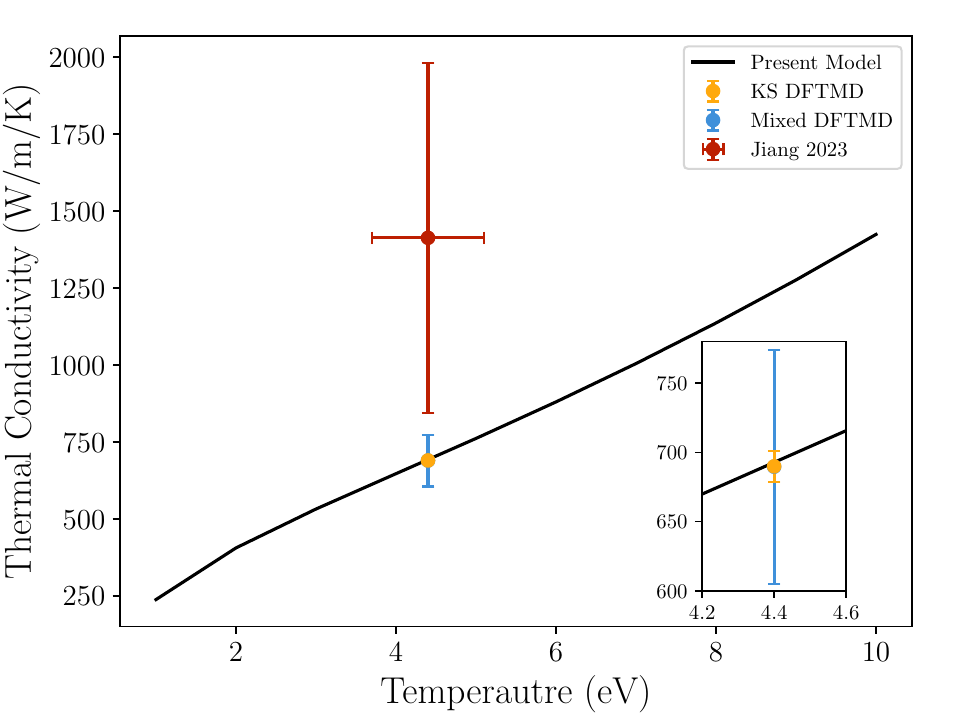}
    \caption{\label{fig:be} Thermal conductivity of beryllium 
    at 1.84~g/cm$^{3}$ and temperatures ranging from 1~eV to 10~eV. These are compared to 
    recent experiment~\cite{JiangNC2023} and two independent DFT-MD methods~\cite{SharmaPOP2026} 
    denoted KS (Kohn-Sham) DFT-MD which is the traditional method and Mixed DFT-MD which is a 
    newer method~\cite{WhitePRL2020}. The inset plot is the region around the DFT-MD 
    values zoomed in so that the error bars on the DFT-MD data can be seen.}
\end{figure}

Recent studies in laser-based ICF have focused on alternative ablator materials, of which 
beryllium has been proposed to replace the current use of high density carbon~\cite{KlinePOP2016, LandenPOP2024}.
Additionally, beryllium is used as a liner material in Magnetic Liner Inertial Fusion 
(MagLIF)~\cite{GomezPRL2014}. Both of these use cases make the thermal conductivity of 
beryllium an important quantity to know in the warm dense matter regime.
As an implosion on either system occurs, the ablator (or liner) will traverse through 
warm dense matter on the way to becoming a plasma. To address this interest, there have been 
experiments~\cite{JiangNC2023} and recent DFT-MD simulations~\cite{SharmaPOP2026} 
which have measured and calculated the thermal conductivity of beryllium at 4.4~eV
and 1.84~g/cm$^{3}$. A comparison with predictions of  the present model can be seen in Fig.~\ref{fig:be},
and due to the reduced computational expense of this method when compared to DFT-MD,
values were calculated from 1~eV to 10~eV. 

The DFT-MD simulations were performed under two frameworks. The first is the traditional Kohn-Sham
DFT, whereas the second uses mixed stochastic 
deterministic DFT. This second framework splits the energy spectrum into low energy
parts which are handled by Kohn-Sham DFT and high energy parts which use stochastic
DFT~\cite{WhitePRL2020}. The computational advantage of this method is that convergence at higher 
energies favors stochastic DFT but lower energies favor Kohn-Sham DFT. Both methods
provide excellent agreement with the prediction of the present model, but all theoretical
predictions fall outside the errorbars of the experimental point. This trend falls 
in line with other calculations which have been compared to the experiential data~\cite{JiangNC2023}.

It is worth noting that for the range of conditions in Fig.~\ref{fig:be}, $\Theta < 0.5$, so
it is expected that electron-electron contributions will be fairly small. 
Furthermore, the higher average charge state of beryllium weakens the importance of electron-electron interactions in comparison to hydrogen at the same $\Theta$ value. 
Due to
this, values for the Lorentz plasma model are nearly identical to the prediction of the full system at these conditions. 
Therefore, the good agreement with the DFT-MD simulation results is also expected. 

\section{Conclusion}
\label{sec:conc}

A model for electronic transport coefficients spanning from traditional plasma
physics into the warm dense matter regime has been developed. This model combines
the Boltzmann-Uehling-Uhlenbeck equation with the potential of mean force to 
model both the degeneracy and correlations present in warm dense matter. Results are
compared to DFT-MD for hydrogen, carbon, and beryllium and display good agreement
at a range of conditions spanning the warm dense matter regime. 

To obtain transport coefficients, the Boltzmann-Uehling-Uhlenbeck kinetic equation 
was solved using the Chapman-Enskog expansion. This expansion involves perturbing
the distribution function from equilibrium and successively solving for the 
perturbation. The perturbation of the distribution function is then connected 
to thermodynamic forces in the system and thermodynamic fluxes are calculated with 
the perturbation. Transport coefficients are the linear relationship between these 
two which are solved for using a polynomial expansion. In the end, to calculate a 
transport coefficient, various ``bracket integrals'' must be solved in a linear
system. 
The calculation presented here extended the Chapman-Enskog solution to the more general BUU equation, which accounts for Fermi statistics, while also containing the traditional classical Boltzmann equation in the classical limit. 

The advantage of the present model is that in the warm dense matter regime, 
accurate tables of transport coefficients can be made in a fraction of the 
time it would take other quantum mechanical based methods such as DFT-MD. Further, it was shown that 
DFT-MD results agree with the results presented here only in the Lorentz plasma limit, which corresponds to turning off electron-electron interactions in the model. 
This further establishes that DFT-MD does not account for electron-electron interactions in the calculation of transport coefficients. 
This is related to the fact that DFT is 
a mean field theory. In regions where it is expected that electron-electron
contributions are small, and the atomic physics is simple, DFT-MD and the present
theory show excellent agreement across a variety of materials. 

This work is a natural extension of mean force kinetic theory~\cite{BaalrudPRL2013, BaalrudPOP2019, DaligaultPRL2016, DaligaultPRL2016, DaligaultPOP2018, ShafferPRE2020_CE, RightleyPRE2021, BabatiPRE2026}
that now provides the ability to calculate all the transport coefficients needed for hydrodynamic simulations spanning plasma and warm dense matter conditions.

\begin{acknowledgments}
The authors would like to thank Dr.~Charles Starrett for access to the ``PYRRHO'' AA-TCP 
code as well as Dr.~Armin Bergermann and Dr.~Mandy Bethkenhagen 
for helpful discussions and providing data to compare with.
This work is funded by the U.S. Department of Energy NNSA Center of Excellence 
under cooperative Agreement No. DE-NA0004146 and by the Department of Energy [National Nuclear Security 
Administration] University of Rochester "National Inertial Confinement Fusion Program" under Award 
Number(s) DE-NA0004144. This report was prepared as an account of work sponsored by an agency the United 
States Government. Neither the United States Government nor any agency thereof, nor any of their 
employees, makes any warranty, express or implied, or assumes any legal liability or responsibility for 
the accuracy, completeness, or usefulness of any information, apparatus, product, or process disclosed, 
or represents that its use would not infringe privately owned rights. Reference herein to any specific 
commercial product, process, or service by trade name, trademark, manufacturer, or otherwise does not 
necessarily constitute or imply its endorsement, recommendation, or favoring by the United States 
Government or any agency thereof. The views and opinions of authors expressed herein do not necessarily 
state or reflect those of the United States Government or any agency thereof.
\end{acknowledgments}

\appendix

\section{Reduction of  Bracket Integrals}
\label{sec:brackets}
To calculate the transport coefficients in Sec.~\ref{sec:electrons}, the $\Lambda_{ee}^{pq}$ 
integrals must be calculated using Eq.~(\ref{eqn:lambda_int}), which in turn 
depend on bracket integrals. The challenge with the choice of the quantum Sonine 
polynomials is that they become extremely complicated expressions at even degree 2. 
This makes calculating bracket integrals in terms of them difficult. Luckily they
are still polynomials, and since bracket integrals are bilinear, they can be expanded
in terms of Sonine polynomial coefficients. Let 
\begin{equation}
    \mathcal{L}^{(q)}_{3/2} \qty(\mathcal{P}^2) = \sum_{i=0}^{q} L_{i,3/2}^{(q)} \mathcal{P}^{2i},
\end{equation}
where $L_{i,3/2}^{(q)}$ is the polynomial coefficient for the $i$th term of the $q$-degree
quantum Sonine polynomial. All bracket integrals can then be decomposed to 
\begin{align}
    \Big[\mathcal{L}^{(q)}_{3/2} \qty(\mathcal{P}^2) \bm{\mathcal{P}}&, \mathcal{L}^{(p)}_{3/2} \qty(\mathcal{P}^2) \bm{\mathcal{P}}\Big] \nonumber \\ 
    &=\sum_{i=0}^{q}\sum_{j=0}^{p} L_{i,3/2}^{(q)}L_{j,3/2}^{(p)} \qty[ \mathcal{P}^{2i}\bm{\mathcal{P}}, \mathcal{P}^{2j}\bm{\mathcal{P}}].
\end{align}
This is simply a change of basis procedure, but reducing the bracket integrals in
this way, and then reconstructing the quantum Sonine polynomials
is much easier than direct calculation. 

Starting with the first term in Eq.~(\ref{eqn:lambda_int}), and writing it out
in the new basis using the definition, Eq.~(\ref{eqn:brack_p}),
\begin{align}
    \Big[\mathcal{P}^{2p} \bm{\mathcal{P}}, \mathcal{P}^{2q} \bm{\mathcal{P}}\Big]_{ei}^{\prime} 
         &= \frac{1}{n_e n_i} \int d^3p_i d^3p_e d\Omega \frac{d\sigma}{d\Omega} h_e h_i u \nonumber \\
         &\times \mathcal{P}_e^{2p}\bm{\mathcal{P}}_e \cdot \qty[\mathcal{P}_e^{2q}\bm{\mathcal{P}}_e - \hat{\mathcal{P}}_e^{2q}\bm{\hat{\mathcal{P}}}_e].
\end{align}
Recall that the index $i$ denotes classical ions, and index $e$ denotes quantum mechanical 
electrons. This means $h_i = f^{(0)}_i$ whereas $h_e = f^{(0)}_e \qty(1 - \theta_e \hat{f}^{(0)}_e)$.
Further, since ions are classical, the limit $\beta\mu \to -\infty$ is taken, which limits to the Maxwellian distribution for ions 
\begin{equation}
    f^{(0)}_i = \frac{n_i}{\pi^{3/2} p_{ti}^{3/2}} \exp\qty(-\frac{p_i^2}{p_{ti}^2}),
\end{equation}
where $p_{ti} = \sqrt{2m_i\kb T}$ is the thermal momentum. The electronic distribution
function is still given by Eq.~(\ref{eqn:FD}) with $\delta_i = -1$. The reduction
of the integral from this point is similar to the process explained in Ref.~\cite{BabatiPRE2026}.
The integration coordinates are changed from the species momenta to a normalized relative velocity
and normalized ion velocity basis defined as 
\begin{subequations}
\begin{equation}
    \bm{u} = \sqrt{\frac{m_e}{2 \kb T}}\qty(\frac{\bm{p}_i}{m_i}-\frac{\bm{p}_e}{m_e})
\end{equation}
and
\begin{equation}
    \bm{v}_i = \sqrt{\frac{m_e}{2\kb T}}\frac{\bm{p}_i}{m_i}.
\end{equation}
\end{subequations}
In these coordinates,
the mass ratio expansion can be taken, $m_e/m_i \to 0$ which means the ionic
distribution function becomes a Dirac delta function centered at $\bm{v}_i = 0$.
From this point, it is a matter of performing as many integrals as possible analytically, yielding 
\begin{align}
 \label{eqn:ie_bracket_int}
    \Big[&\mathcal{P}^{2p} \bm{\mathcal{P}}, \mathcal{P}^{2q} \bm{\mathcal{P}}\Big]_{ei}^{\prime} 
         = \frac{4}{\sqrt{\pi}c_{\Theta,e}^{p+q+1}\mathcal{Q}_{1/2}(\beta\mu_e)} \\ \nonumber &\times \sqrt{\frac{2\kb T}{m_e}} \int_{0}^{\infty} du \frac{e^{u^2 - \beta\mu_e}}{(e^{u^2 - \beta\mu_e}+1)^2} 
          u^{2(p+q+2)+1} \sigma_{p}.
\end{align} 
Here, $\sigma_p$ is the momentum transfer cross section defined as 
\begin{equation}
    \sigma_p = \int d\Omega \frac{d\sigma}{d\Omega} (1-\cos\theta)
    \label{eqn:mom_trans}
\end{equation}
where $\theta$ is the scattering angle of a collision.

Equation~(\ref{eqn:ie_bracket_int}) contains the ion-electron interactions within the system.
The rest of the bracket integrals in Eq.~(\ref{eqn:lambda_int}) can be combined into a single expression which 
resembles a total bracket integral for a single species, which in this case is the 
electrons. Representing the bracket integral in a form similar to Eq.~(\ref{eqn:bracket_other}) is helpful
so that the integral calculated is
\begin{align}
    \Big[ &\mathcal{P}^{2p}\bm{\mathcal{P}} , \mathcal{P}^{2q}\bm{\mathcal{P}}\Big] = \frac{1}{4n^2}\int d^3{p_\alpha}d^3{p_\beta}d\Omega \frac{d\sigma}{d\Omega} u h_\alpha h_\beta \nonumber\\ 
    \times & \Big[ \mathcal{P}_\alpha^{2p} \bm{\mathcal{P}}_\alpha + \mathcal{P}_\beta^{2p} \bm{\mathcal{P}}_\beta - \hat{\mathcal{P}}_\alpha^{2p} \bm{\hat{\mathcal{P}}}_\alpha - \hat{\mathcal{P}}_\beta^{2p} \bm{\hat{\mathcal{P}}}_\beta \Big] \nonumber \\
    \cdot & \Big[ \mathcal{P}_\alpha^{2q} \bm{\mathcal{P}}_\alpha + \mathcal{P}_\beta^{2q} \bm{\mathcal{P}}_\beta - \hat{\mathcal{P}}_\alpha^{2q} \bm{\hat{\mathcal{P}}}_\alpha - \hat{\mathcal{P}}_\beta^{2q} \bm{\hat{\mathcal{P}}}_\beta \Big]
    \label{eqn:electron_bracket}
\end{align}
where the indices here denote different species. In the case we consider
both species $\alpha$ and $\beta$ are electrons, so the differential scattering cross section
must be calculated for electron-electron collisions instead of ion-electron collisions.
In this form, it is clear to see that if $p=0$ or $q=0$ 
the integral will be identically 0 by conservation of momentum. It is also symmetric, so for the expressions in 
Sec.~\ref{sec:electrons} only three integrals are required, $(p,q)=(1,1), (1,2), (2,2)$.

To reduce this integral it is convenient to use the normalized center of mass and relative 
momentum frame 
\begin{subequations}
\begin{equation}
    \bm{G} = \frac{1}{\sqrt{2m_{e}\kb T}} \qty(\frac{\bm{P}_\alpha + \bm{P}_\beta}{2}),
\end{equation}
\begin{equation}
    \bm{g} = \frac{1}{\sqrt{2m_e \kb T}} \qty(\bm{P}_\alpha - \bm{P}_\beta).
\end{equation}
\end{subequations}
Note, the jacobian for this transformation is $d^3 p_\alpha d^3 p_\beta = (2m_{e} \kb T)^3d^3G d^3g$.
Substituting this transformation into the integral and reducing gives 
\begin{align}
    \Big[ &\mathcal{P}^{2p}\bm{\mathcal{P}} , \mathcal{P}^{2q}\bm{\mathcal{P}}\Big] = \frac{1}{4 \pi^3 c_{\Theta, e}^{q+p+1}\mathcal{Q}_{1/2}^2\qty(\beta\mu_e)}\sqrt{\frac{2\kb T}{m_e}} \nonumber \\
    \times & \int d^3 g d\Omega \frac{d\sigma}{d\Omega} g f_{\textrm{eff}}(g, \theta)
\end{align}
where $f_\textrm{eff}(g, \theta)$ is an effective distribution defined as 
\begin{align}
    f_{\textrm{eff}}(g, \theta) = \int d^3 G h_\alpha h_\beta F_{pq}\qty(\bm{G},\bm{g},\hat{\bm{g}}),
    \label{eqn:effective_dist}
\end{align}
where $F_{pq}\qty(\bm{G},\bm{g},\hat{\bm{g}})$ corresponds 
to the polynomial in Eq.~(\ref{eqn:electron_bracket}). For the integrals of interest
they are given by
\begin{subequations}
\begin{align}
    F_{11}&\qty(\bm{G},\bm{g},\hat{\bm{g}}) =
    \qty(\qty(\Gg)^2 \bm{g} - \qty(\Ggh) \hat{\bm{g}})^2, \\
    F_{12}&\qty(\bm{G},\bm{g},\hat{\bm{g}}) = 2\qty(\qty(\Gg)^2 - \qty(\Ggh)^2)^2 \nonumber \\
          &+ 2\qty(G^2 + \frac{g^2}{4})g^2\qty(\qty(\Gg)^2 \bm{g} - \qty(\Ggh) \hat{\bm{g}})^2,\\
    F_{22}&\qty(\bm{G},\bm{g},\hat{\bm{g}}) = \qty(12G^2 + 2g^2)\qty(\qty(\Gg)^2 - \qty(\Ggh)^2)^2 \nonumber \\
          &+ 4\qty(G^2 + \frac{g^2}{4})^2\qty(\qty(\Gg)^2 \bm{g} - \qty(\Ggh) \hat{\bm{g}})^2.
\end{align}
\end{subequations}
The product of distribution functions has a convenient form in these coordinates, given by 
\begin{equation}
    h_\alpha h_\beta = \frac{1}{4\qty(\cosh\qty(a) + \cosh\qty(b))\qty(\cosh\qty(a)+\cosh\qty(\hat{b}))}
\end{equation}
where $a = G^2 + g^2/4 - \beta\mu_e$, $b=\bm{G}\cdot\bm{g}$, and $\hat{b}=\hat{\bm{G}}\cdot\hat{\bm{g}}$. 
Note that the center of mass momentum of a binary collision is conserved, so 
$\bm{G} = \hat{\bm{G}}$ and then for elastic collisions,
$g = \hat{g}$, and $\bm{g}\cdot\hat{\bm{g}} = g^2 \cos(\theta)$,
where $\theta$ is the scattering angle.

The simplest way to evaluate these integrals is to use a spherical coordinate 
system where the vector $\bm{g}$ is located along the z axis. This means
$\hat{\bm{g}} = g\qty[\sin\theta\cos\phi \hat{x} + \sin\theta\sin\phi \hat{y} + \cos\theta\hat{z}]$,
where $\theta$ and $\phi$ are scattering angles, and $\hat{x}$, $\hat{y}$, $\hat{z}$ are 
cartesian unit vectors. Similarly, the center of mass momentum vector is given 
in spherical coordinates as $\bm{G} = G\qty[\sin\theta^\prime\cos\phi^\prime \hat{x} + \sin\theta^\prime\sin\phi^\prime \hat{y} + \cos\theta^\prime\hat{z}]$
where $\theta^\prime$ and $\phi^\prime$ are spherical angles. These integrals are 
not as simple as Eq.~(\ref{eqn:ie_bracket_int}), but with these substitutions can 
be calculated numerically.

\section{Quantum Scattering}
\label{sec:scattering}
The general scheme used for quantum scattering is based on the partial wave expansion
of the quantum scattering amplitude~\cite{Sakurai}
\begin{equation}
    f(\theta) = \sum_{l=0}^{\infty}\qty(2l+1)\qty(\frac{e^{i\delta_l}\sin\delta_l}{k})P_l\qty(\cos\theta),
\end{equation}
where $\theta$ is the scattering angle, $l$ is the angular momentum quantum number,
$\delta_l$ is known as the phase shift, $\hbar k = \sqrt{2 m_{\alpha\beta} E}$ defines the $k$
vector of the collision, $m_{\alpha\beta}$ is the reduced mass of a collision, and $P_l(x)$ is 
a Legendre polynomial. The phase shifts $\delta_l$ are determined using the 
variable phase equation~\cite{Calogero}
\begin{align}
    \frac{d\delta_l}{dr}&(r) \nonumber \\
                        &= -\frac{2m_{\alpha\beta}}{\hbar^2 k} V(r) \qty(\hat{j}_l(kr)\cos\qty(\delta_l(r)) - \hat{n}_l(kr)\sin\qty(\delta_l(r)))^2
                        \label{eqn:phase_eqn}
\end{align}
where $\delta_l(r)$ is the phase accumulated by a spherical wave with angular momentum
number $l$ a distance $r$ away from the scattering center, $V(r)$ is the central 
scattering potential, and $\hat{j}_l(kr)$ and $\hat{n}_l(kr)$ are Riccati-Bessel
functions defined as $\hat{j}_l(kr) = krj_l(kr)$ and $\hat{n}_l(kr) = krn_l(kr)$
where $j_l(kr)$ and $n_l(kr)$ are typical spherical Bessel functions. Typically,
Eq.~(\ref{eqn:phase_eqn}) is a stiff differential equation, thus needs an appropriate
solver, LSODA~\cite{LSODA} has proven to be a good choice. For most temperatures, 
the number of partial waves becomes large, and the numerical integration can be 
become unstable, so a Born approximation to the phase shifts is taken according 
to a scheme explained in Ref.~\cite{LinPPCF2023}. The phase shifts are then 
given by 
\begin{equation}
    \delta_l = -\frac{2m_{\alpha\beta}}{\hbar^2 k}\int_0^\infty dr V(r) \qty(\hat{j}_l(kr))^2.
\end{equation}
This form can be obtained by taking the phase shifts in Eq.~(\ref{eqn:phase_eqn})
to be small and the switch between the two methods occurs when this is the case.

At even higher energies a full Born approximation can be taken so that~\cite{Sakurai}
\begin{equation}
    f(\theta) = -\frac{2m_{\alpha\beta}}{\hbar^2 q} \int_0^\infty dr rV(r) \sin\qty(qr),
\end{equation}
where $q = 2k \sin\qty(\theta/2)$. The transition to this description occurs 
when $\max\qty(rV(r)) < \gamma \hbar^2 k^2 /(2m_{\alpha\beta})$ and $\gamma$ is usually
set to 0.07. This is a high energy approximation to the scattering.

The differential cross section used depends on the particles participating 
in the collision. For electrons and ions, after a mass ratio expansion is taken,
the system reduces to the single electron scattering off an unmoving ion. This 
is the textbook case, so $m_{\alpha\beta} = m_e$ and the differential scattering cross section
is given by~\cite{Sakurai}
\begin{equation}
    \frac{d\sigma}{d\Omega} = \qty| f(\theta) |^2.
\end{equation}
Then if the partial wave expansion is taken and substituted into the momentum 
transfer cross section defined in Eq.~(\ref{eqn:mom_trans}) it becomes
\begin{equation}
    \sigma_p = \frac{4\pi}{k^2}\sum_{l=0}^\infty (l+1) \sin^2\qty(\delta_{l+1} - \delta_l).
\end{equation}
For electron-electron collisions, the system has a reduced mass of $m_{\alpha\beta} = m_e/2$
and identical particle effects must be considered. Mainly, after a scattering event,
an electron scattering by an angle of $\theta$ and $\pi-\theta$ are indistinguishable.
Additionally, the total wavefunction of both electrons must be antisymmetric about 
exchange of the particles, so depending on the total spin, the spatial part of the 
wave function is either symmetric or antisymmetric. This means for an ensemble 
of electrons with randomized spins the differential scattering cross section is given by 
\cite{Sakurai}
\begin{equation}
    \frac{d\sigma}{d\Omega} = \frac{1}{4}\qty|f(\theta) + f(\pi-\theta)|^2 + \frac{3}{4}\qty|f(\theta) - f(\pi-\theta)|^2.
\end{equation}
The factors of 1/4 and 3/4 come from the fact that there is one antisymmetric total
spin state two electrons can have (so one symmetric spatial wave function), but three symmetric total spin
states (so three antisymmetric spatial wave functions). The scattering amplitude, $f(\theta)$
is related to the spatial part of the two electron wave function.

\bibliography{refs}

\end{document}